\documentclass{aa}  

\usepackage{graphicx}
\usepackage{txfonts}
\usepackage{makecell}
\usepackage{multirow}
\usepackage{amssymb}
\usepackage{amsmath}
\usepackage{xcolor}
\usepackage[normalem]{ulem}

\usepackage{hyperref}

\hypersetup{colorlinks=true, citecolor=blue, linkcolor=blue}
\begin{document} 

   \titlerunning{The size-luminosity relation from interferometric broad-line region observations}
   \title{The size-luminosity relation of local active galactic nuclei from interferometric observations of the broad-line region}

    \authorrunning{GRAVITY Collaboration}
\author{GRAVITY Collaboration\thanks{GRAVITY is developed in a collaboration by 
the Max Planck Institute for Extraterrestrial Physics, LESIA of Observatoire 
de Paris/Universit\'e PSL/CNRS/Sorbonne Universit\'e/Universit\'e de Paris and 
IPAG of Universit\'e Grenoble Alpes /CNRS, the Max Planck Institute for Astronomy, 
the University of Cologne, the CENTRA - Centro de Astrofisica e Gravita\c c\~ao, 
and the European Southern Observatory.}:
A.~Amorim\inst{1,2}
\and G.~Bourdarot\inst{3}
\and W.~Brandner\inst{4} 
\and Y.~Cao\inst{3} 
\and Y.~Cl\'enet\inst{5} 
\and R.~Davies\inst{3}
\and P.~T.~de~Zeeuw\inst{6} 
\and J.~Dexter\inst{3,7}
\and A.~Drescher\inst{3}  
\and A.~Eckart\inst{8,9} 
\and F.~Eisenhauer\inst{3} 
\and M.~Fabricius\inst{3}
\and H.~Feuchtgruber\inst{3}
\and N.~M.~F\"orster~Schreiber\inst{3} 
\and P.~J.~V.~Garcia\inst{2,10,11} 
\and R.~Genzel\inst{3,12} 
\and S.~Gillessen\inst{3} 
\and D.~Gratadour\inst{5,13} 
\and S.~H\"onig\inst{14}
\and M.~Kishimoto\inst{15} 
\and S.~Lacour\inst{5,16} 
\and D.~Lutz\inst{3} 
\and F.~Millour\inst{17}  
\and H.~Netzer\inst{18} 
\and T.~Ott\inst{3} 
\and T.~Paumard\inst{5} 
\and K.~Perraut\inst{19} 
\and G.~Perrin\inst{5}
\and B.~M.~Peterson\inst{20}
\and P.~O.~Petrucci\inst{19} 
\and O.~Pfuhl\inst{16}
\and M.~A.~Prieto\inst{21}  
\and S.~Rabien\inst{3}
\and D.~Rouan\inst{5}
\and D.~J.~D.~Santos\inst{3}\thanks{Corresponding author: D. J. D. Santos (dsantos@mpe.mpg.de)}
\and J.~Shangguan\inst{3}
\and T.~Shimizu\inst{3}
\and A.~Sternberg\inst{18,22} 
\and C.~Straubmeier\inst{8} 
\and E.~Sturm\inst{3} 
\and L.~J.~Tacconi\inst{3} 
\and K.~R.~W.~Tristram\inst{10}  
\and F.~Widmann\inst{3} 
\and J.~Woillez\inst{16}}

\institute{Universidade de Lisboa - Faculdade de Ci\^{e}ncias, Campo Grande, 
1749-016 Lisboa, Portugal
\and CENTRA - Centro de Astrof\'isica e Gravita\c{c}\~{a}o, IST, Universidade de Lisboa, 
1049-001 Lisboa, Portugal
\and Max Planck Institute for Extraterrestrial Physics (MPE), Giessenbachstr.1, 
85748 Garching, Germany
\and Max Planck Institute for Astronomy, K\"onigstuhl 17, 69117, Heidelberg, Germany
\and LESIA, Observatoire de Paris, Universit\'e PSL, CNRS, 
Sorbonne Universit\'e, Univ. Paris Diderot, Sorbonne Paris Cit\'e, 5 place Jules Janssen, 
92195 Meudon, France
\and Leiden University, 2311EZ Leiden, The Netherlands
\and Department of Astrophysical \& Planetary Sciences, JILA, University of Colorado, 
Duane Physics Bldg., 2000 Colorado Ave, Boulder, CO 80309, USA
\and I. Institute of Physics, University of Cologne, Z\"ulpicher Stra{\ss}e 77, 
50937 Cologne, Germany
\and Max Planck Institute for Radio Astronomy, Auf dem H\"ugel 69, 53121 Bonn, Germany
\and European Southern Observatory, Alonso de C\'ordova 3107, Casilla 19001, Vitacura, Santiago, Chile
\and Faculdade de Engenharia, Universidade do Porto, rua Dr. Roberto Frias, 
4200-465 Porto, Portugal
\and Departments of Physics and Astronomy, Le Conte Hall, University of California, 
Berkeley, CA 94720, USA
\and Research School of Astronomy and Astrophysics, Australian National University, 
Canberra, ACT 2611, Australia
\and Department of Physics and Astronomy, University of Southampton, Southampton, UK
\and Department of Physics, Kyoto Sangyo University, Kita-ku, Japan
\and European Southern Observatory, Karl-Schwarzschild-Str. 2, 85748 Garching, Germany
\and Universit\'e C\^ote d'Azur, Observatoire de la C\^ote d'Azur, CNRS, 
Laboratoire Lagrange, Nice, France
\and School of Physics and Astronomy, Tel Aviv University, Tel Aviv 69978, Israel
\and Univ. Grenoble Alpes, CNRS, IPAG, 38000 Grenoble, France
\and Retired
\and Instituto de Astrof\'isica de Canarias (IAC), E-38205 La Laguna, Tenerife, Spain
\and Center for Computational Astrophysics, Flatiron Institute, 162 5th Ave., 
New York, NY 10010, USA
}

   \date{Received xxxx xx, 2023; accepted xxxx xx, 2023}

 
  \abstract{By using the GRAVITY instrument with the near-infrared (NIR) Very Large Telescope Interferometer (VLTI), the structure of the broad (emission-)line region (BLR) in active galactic nuclei (AGNs) can be spatially resolved, allowing the central black hole (BH) mass to be determined. This work reports new NIR VLTI/GRAVITY interferometric spectra for four type 1 AGNs (Mrk 509, PDS 456, Mrk 1239, and IC 4329A) with resolved broad-line emission. Dynamical modelling of interferometric data constrains the BLR radius and central BH mass measurements for our targets and reveals outflow-dominated BLRs for Mrk 509 and PDS 456. We present an updated radius-luminosity (R-L) relation independent of that derived with reverberation mapping (RM) measurements using all the GRAVITY-observed AGNs. We find our R-L relation to be largely consistent with that derived from RM measurements except at high luminosity, where BLR radii seem to be smaller than predicted. This is consistent with RM-based claims that high Eddington ratio AGNs show consistently smaller BLR sizes. The BH masses of our targets are also consistent with the standard $M_\mathrm{BH}$-$\sigma_*$ relation. Model-independent photocentre fitting shows spatial offsets between the hot dust continuum and the BLR photocentres (ranging from $\sim$17 $\mu$as to 140 $\mu$as) that are generally perpendicular to the alignment of the red- and blueshifted BLR photocentres. These offsets are found to be related to the AGN luminosity and could be caused by asymmetric K-band emission of the hot dust, shifting the dust photocentre. We discuss various possible scenarios that can explain this phenomenon.}

   \maketitle
%

\section{Introduction}

Many earlier works have shown that supermassive black holes (SMBHs) reside in the centre of most galaxies \citep[e.g.][]{Soltan1982, Rees1984, Ferrarese2005, Davis2014, Padovani2017}. The $M_\mathrm{BH}$-$\sigma_*$ relation (the relationship between the central BH mass and the stellar velocity dispersion of the host galaxy) provides strong evidence for SMBH-host galaxy coevolution \citep[e.g.][]{Magorrian1998, Gebhardt2000a, Treu2004, Kormendy2013, Caglar2020}. This requires robust measurements of BH masses. There are many ways of doing this, such as spatially resolving stellar kinematics \citep[e.g.][]{Gebhardt2000a, Sharma2014}, measuring the kinematics of (ionised) gas \citep[e.g.][]{Davies2004a, Davies2004b, Hicks2008, Davis2013}, utilising megamaser kinematics with very long baseline interferometry (VLBI) \citep[e.g.][]{Kuo2020, Wagner2013, vandenBosch2016}, and reverberation mapping (RM) \citep[e.g.][]{Peterson1993, Lira2018, Cackett2021}. Other indirect methods involve scaling relations, relying on single-epoch optical/near-infrared (NIR) spectroscopy to observe several line estimators and to measure their fluxes and widths (e.g. H$\alpha$, H$\beta$, Mg {\tiny II} $\lambda$2798, and C {\tiny IV} $\lambda$1549) \citep[e.g.][]{Kaspi2000, Greene2005, Shen2012} as well as coronal lines to estimate the accretion disc temperature  \citep[e.g.][]{Prieto2022}.

When the SMBH is 'active', it accretes surrounding material and forms an accretion disc. The brightness of the resulting active galactic nucleus (AGN) dilutes the signal from the stars and gas surrounding it (in addition to the gas being subject to other forces than gravity, e.g. radiation pressure, winds, etc.), hindering BH mass measurements via stellar and gas kinematics \citep{Gebhardt2000a}. The intense radiation can trigger megamasers and thus allow for this SMBH mass measurement method, but the need for a nearly edge-on line of sight (LOS) allows for only type 2 AGNs \citep{vandenBosch2016} to be observed.
    
Reverberation mapping campaigns utilise the fluctuations over time of the continuum flux originating from the accretion disc \citep{Blandford1982, Peterson1993, Peterson2004}. These variations are then echoed or reverberated in the surrounding gas of the broad-line region (BLR). The time lag between the continuum and BLR variations can be measured from their light curves and utilised to estimate the radius of the BLR under the assumption that the measured time delay represents the light-crossing time between the accretion disc and BLR. The mass of the central BH is then calculated via the following formula:
\begin{equation}
    \label{eqn:virial_mass}
    M_{\rm BH} = \frac{f{R}_{\rm BLR}v^2}{G},
\end{equation}
where $G$ is the gravitational constant, $M_\mathrm{BH}$ is the central BH mass, $R_\mathrm{BLR}$ is the BLR radius, and $v$ is the full width at half maximum FWHM or dispersion ($\sigma$) of the line observed with RM. We note that $f$ is the virial factor, which accounts for the unknown geometry and structure of the BLR. 

Previous works have revealed that local AGNs exhibit a relationship between the BLR radius measured from AGN emission lines (e.g. H$\alpha$ and H$\beta$) and AGN continuum luminosity (usually at 5100 $\AA$), the so-called R-L relation \citep{Kaspi2000, Bentz2013}. The R-L relation, combined with a measurement of the gas velocity (e.g. the FWHM of the broad emission lines), can then be used to estimate BH masses of distant AGNs using Eq.~\ref{eqn:virial_mass} \citep[e.g.][]{Laor1998, Vestergaard2006, Shen2023}. However, BH mass estimations via single-epoch spectroscopy are only accurate if the R-L relation accurately represents the whole AGN population. Hence, confirming the relation with a more diverse and broad AGN sample is important. The accretion rate or $L$/$L_\mathrm{edd}$ has been suggested to serve as a 'third' parameter in the R-L relation, as highly accreting AGNs seem to have smaller BLR sizes than what is predicted by the canonical R-L relation \citep{Du2014b, Du2015, Du2016, Du2019}. Most of these highly accreting sources are also highly luminous (log $\lambda L_{\lambda} (5100 \AA)$/erg s$^{-1}$ $\gtrsim$ 44; \citealt{Du2018}), and hence the deviation from the canonical R-L relation is more prominent at higher luminosities. \citet{Du2018} and \citet{Du2019} show that using  $\mathcal{R}$(Fe {\tiny II}) (the equivalent width ratio of Fe {\tiny II} and H$\beta$) as a proxy of accretion rate can reduce the scatter of the R-L relation. Not accounting for the accretion rate could then lead to BH mass estimations via single-epoch spectroscopy and the R-L relation to be overestimated by as much as $\sim$ 0.3 dex \citep{Alvarez2020, GRAVITY2023a}.
    
Reverberation mapping measurements have concluded that the BLR gas is mostly dominated by Keplerian motion and is therefore orbiting around the central BH \citep[e.g.][]{Gaskell2000, Denney2009, Bentz2010a, Du2016, Grier2017a, Williams2018, Bentz2021b, Villafana2023}. The BLR could also have considerable turbulence in the direction perpendicular to the BLR mid-plane, with significant inflow and outflow velocity components \citep{Osterbrock1978, Ulrich1996}. Several pieces of evidence of possible inflow and outflow signatures for some BLRs have been found \citep[e.g.,][]{Kollatschny1997, Denney2009, Bentz2009, Bentz2010a, Du2016}. On the other hand, three BLRs whose kinematics were resolved by GRAVITY were found to be rotation dominated \citep{GRAVITY2018, GRAVITY2020a, GRAVITY2021a}. Hence, a single description of the BLR has not been established yet.
    
A crucial limitation of RM is the uncertain virial factor,  $f$. Usually, $f$ is calibrated assuming that AGNs follow a similar $M_\mathrm{BH}$-$\sigma_*$ relation with those of quiescent galaxies. Although $f$ is different for each AGN and cannot be determined without velocity-resolved RM data, the average $f$ can be calculated for a sample of AGNs. The average virial factor, $\langle f \rangle$, serves as a scaling factor of RM-derived BH masses to match the said relation. Various works have different values of $\langle f \rangle$ \citep[e.g.][]{Onken2004, Woo2010, Graham2011, Park2012, Batiste2017}. Source variability and line width uncertainties could also lead to uncertainties in the measured time lag (see \citealt{Vestergaard2011} and references therein). RM campaigns are also usually limited by the luminosity of the AGNs they can target, as more luminous targets should require much longer RM campaigns to measure the expected time lags. Recent RM surveys though are working to expand the high luminosity range \citep[e.g.][]{Woo2023}.

Spatially resolving the BLR was initially difficult to accomplish due to its very small angular size ($\lesssim$ 10$^{-4}$ arcseconds; \citealt{Blandford1982}). However, thanks to GRAVITY, the second-generation NIR beam combiner in the Very Large Telescope Interferometer (VLTI), the sensitivity of NIR interferometry has been exceptionally improved, and the light received by all four 8m unit telescopes (UTs) has been combined to yield six simultaneous baselines \citep{GRAVITY2017}. This has allowed us to not just spatially resolve the BLR, but also obtain measurements of the SMBH mass and investigate the BLR structure at a high-angular resolution. To this end, we carried out an ESO Large Programme to observe the brightest type 1 AGNs, which span four orders of magnitude in luminosity, to spatially resolve their BLR and measure their central BH masses. This programme also aims to further understand the BLR structure and its intricacies, such as inflow and outflow motions that may be present in some systems, and to investigate GRAVITY-derived BLR radius and BH mass measurements and how they compare with scaling relations of local AGNs, such as the R-L and $M_\mathrm{BH}$-$\sigma_*$ relations. Three BLRs from this sample have already been spatially resolved by GRAVITY \citep{GRAVITY2018, GRAVITY2020a, GRAVITY2021a, GRAVITY2021b}. In parallel, several works have focussed on resolving the dust sublimation region \citep{GRAVITY2020b, GRAVITY2021a}, and using the observed dust size to estimate BLR sizes \citep{GRAVITY2020c, GRAVITY2023b}.
  
In this work, we present an analysis of four new GRAVITY observations of type 1 AGNs, namely Mrk 509, PDS 456, Mrk 1239, and IC 4329A, all in the local Universe ($z$ $<$ 0.2). For the entirety of our paper, we adopt a Lambda Cold Dark Matter ($\Lambda$CDM) cosmology with $\Omega_m$ = 0.308, $\Omega_\Lambda$ = 0.692, and $H_0$ = 67.8 km s$^{-1}$ Mpc$^{-1}$ \citep{Planck2016}. We discuss the targets' properties in Sect. \ref{sec:targets}. We describe the observations, data reduction, and the resulting flux and differential phase spectra in Sect. \ref{sec:obs}.  We report the model-independent photocentre positions of each target's BLR in Sect. \ref{sec:photocentre}. We describe our BLR model in Sect. \ref{sec:modelling}, and discuss the importance of differential phase data in our BLR modelling in Sect. \ref{sec:visphi_importance}. We present the results of our BLR model fitting in Sect. \ref{sec:discussion}, while we place the derived BLR sizes and BH masses of our targets in the context of the R-L and $M_\mathrm{BH}$-$\sigma_*$ relations in Sects. \ref{sec:R-L} and \ref{sec:M-sigma}, respectively. We also observed an offset between the BLR and continuum photocentres of our targets and discuss its possible origin in Sect. \ref{sec:blr_offset_vs_lum}. We also explain how we calculated the virial factors of our targets in Sect. \ref{sec:f}. Finally, we present our conclusions and future prospects in Sect. \ref{sec:conclusion}.

\section{Targets}
\label{sec:targets}

All four targets were observed with GRAVITY as part of a large programme (and initial pilot projects) that aims to spatially resolve the BLR and measure BH masses for a sample of the brightest (nuclear luminosity of $K$ $\lesssim$ 10 mag and $V$ $\lesssim$ 15 mag) type 1 (Sy1 or QSOs with broad Pa$\alpha$ or Br$\gamma$ emission lines) AGNs in the local Universe\footnote{All GRAVITY observations were made using the ESO Telescopes at La Silla Paranal Observatory with programme IDs 1104.C-0651(C), 1103.B-0626(B), 099.B-0606(A), and 0101.B-0255(B).}. Table \ref{tab:targets} summarises the properties of our four targets and the three other targets that are already published.

\begin{table*}
\caption{Physical properties of our four new targets and the three targets that were already observed by GRAVITY.}      
\label{tab:targets}     
\centering                         
\begin{tabular}{ccccccccc}       
\hline              
\hline
Object & \makecell{RA \\ (J2000)} & \makecell{Dec \\ (J2000)} & $z$ & \makecell{log $\lambda L_\lambda$ (5100 $\AA$) \\ (erg s$^{-1}$)} & Ref. & \makecell{$\sigma_*$ \\ (km ${\rm s}^{-1}$)} & Ref. & \makecell{$D_\mathrm{A}$ \\ (Mpc)} \\
(1) & (2) & (3) & (4) & (5) & (6) & (7) & (8) & (9) \\
\hline
Mrk 509 & 20:44:09.738 & $-$10:43:24.54 &  0.0344 & 44.19 & 1 & 182 & 8 & 144 \\
PDS 456 & 17:28:19.796 & $-$14:15:55.87 & 0.185 & 46.30 & 2 & 182$^{1}$ & This work & 657 \\
Mrk 1239 & 09:52:19.102 & $-$01:36:43.46 & 0.020 & 44.40$^{2}$ & 3 & $\sim$250$^{3}$ & 9 & 86 \\ 
IC 4329A & 13:49:19.266 & $-$30:18:33.97 & 0.016 & 43.51 & 4 & $\sim$225 & 9 & 69 \\
\hline
3C 273 & 12:29:06.700 & $+$02:03:08.60 & 0.158 & 45.90 & 1 & 210 & 10 & 582 \\
NGC 3783 & 11:39:01.762 & $-$37:44:19.21 & 0.0097 & 43.02 & 6 & 95 & 11 & 42 \\
IRAS 09149-6206 & 09:16:09.39 & $-$62:19:29.90 & 0.0573 & 44.92d & 5,7 & 250$^{4}$ & 7 & 236 \\
\hline
\end{tabular}
\vspace{0.1mm}
{\footnotesize \begin{itemize}
\item[$^{1}$] The stellar velocity dispersion of PDS 456 is estimated in this work by using the measured dynamical mass of PDS 456 by \citet{Bischetti2019} and virial theorem.
\item[$^{2}$] Although \citet{Pan2021} measures the extinction-corrected optical luminosity of Mrk 1239, caution is still needed as Mrk 1239 is widely known to show high polarisation and high extinction.
\item[$^{3}$] The stellar velocity dispersion of Mrk 1239 is highly uncertain due to the shallowness of its stellar features \citep{Onken2004}
\item[$^{4}$] The dispersion of IRAS 09149-6206 is based on the [O{\small III}] line and is very uncertain \citep{GRAVITY2020a}.
\end{itemize}
\vspace{0.5mm}
\begin{flushleft}
\textbf{Notes:} 
Col.~(1): Object name. 
Col.~(2): Right ascension.
Col.~(3): Declination.
Col.~(4): Redshift from NASA/IPAC Extragalactic Database (NED).
Col.~(5): AGN optical luminosity at 5100~\AA.
Col.~(6): References for $\lambda L_\lambda$ (5100 \AA).
Col.~(7): Stellar velocity dispersion.
Col.~(8): References for $\sigma_*$.
Col.~(9): Angular diameter distance.
\vspace{0.5mm}
\\
\textbf{References:} 
(1) \cite{Du2019}, 
(2) \cite{Nardini2015},
(3) \cite{Pan2021}, 
(4) \cite{Bentz2023}
(5) \cite{Koss2017}, 
(6) \cite{Bentz2021a},
(7) \cite{GRAVITY2020a},
(8) \cite{Grier2013},
(9) \cite{Oliva1999},
(10) \cite{Husemann2019},
(11) \cite{Onken2004}
\end{flushleft}}
\end{table*}

\subsection{Mrk 509}
Mrk 509 is a type 1 Seyfert galaxy that shows significant outflow signatures in terms of mildly relativistic ($\sim$ 0.14-0.2$c$) FeXXVI K$\alpha$ and K$\beta$ absorption features \citep{Cappi2009} detected in its X-ray spectra. These imply a possible varying structure and geometry \citep{Cappi2009}. RM campaigns have estimated the BLR size of Mrk 509 to be $R_\mathrm{BLR}$ $\sim$ 80 ld \citep{Carone1996, Peterson1998, Bentz2009, Shablovinskaya2023}. Its central BH mass was also calculated to be $M_\mathrm{BH} = 10^{7.9} - 10^{8.3} \ M_\odot$ \citep{Peterson1998, Peterson2004}.

\subsection{PDS 456}
PDS 456 is known to be the most luminous radio-quiet AGN in the local Universe ($z$ $\lesssim$ 0.3) \citep{Torres1997, Simpson1999, Bischetti2019, Yun2004}. Aside from its brightness and proximity, it has been extensively studied due to its ultra-fast outflows (UFOs) of highly ionised Fe \citep{Reeves2003}. These outflows possess high kinetic power ($\sim$20\% of its bolometric luminosity; \citealt{Luminari2018}) and are also radiatively driven, which can be explained by the AGN's high Eddington ratio \citep{Matzeu2017}. Its BH mass, unfortunately, is not well-studied. No RM measurements are available for PDS 456 because such a luminous target should produce very long time lags ($\sim$ several years), requiring decade-long RM campaigns. Therefore, BH mass estimations based on empirical relations from RM are used \citep{Reeves2009, Nardini2015}. Its central BH mass is estimated to be $M_\mathrm{BH} \sim 10^{9} \ M_\odot$. Its BLR size is also implied to be large; \citet{GRAVITY2020c} partially resolved the continuum hot dust emission region of PDS 456 and used the measured continuum size from the fringe tracker data and the differential visibility amplitude data from the science channel (SC) to indirectly infer its BLR size. However, their estimate is highly uncertain due to the weak correlation between the SC differential amplitude and the BLR size for very small angular sizes relative to the baseline resolution. \citet{GRAVITY2023a} estimated its BH mass and BLR radius to be $M_\mathrm{BH} \sim 10^{8.68} \ M_\odot$ and $R_\mathrm{BLR}$ $\sim$ 150 ld from the relation between the BLR and dust continuum size. It is interesting to note that \citet{GRAVITY2023a} measured a smaller BH mass than previous works have estimated, and this deviation is suggested as a consequence of its high Eddington ratio \citep{Du2019} that has been implied by its strong Fe{\small II} features and lack of [O{\small III}] emission lines \citep[e.g.,][]{Simpson1999}.

\subsection{Mrk 1239}
Mrk 1239 is a narrow-line Seyfert 1 (NLSy1) galaxy well known for its high polarisation \citep{Goodrich1989, Smith2004}, relatively redder optical-IR colour compared to other typical NLSy1s, and presence of a radio jet-like structure \citep{Orienti2010, Doi2015} oriented perpendicular to the polarisation angle. Mrk 1239 has not been a target for RM campaigns, but previous works have estimated its BH mass and BLR size via indirect estimates from various scaling relations. They all point to the same conclusion that Mrk 1239 has a small but uncertain BH mass, $M_\mathrm{BH} \sim 10^{5.7} - 10^{7.0} \ M_\odot$ \citep[e.g.][]{Kaspi2005, Greene2005, Ryan2007, Du2014a, Buhariwalla2020, Pan2021, GRAVITY2023b}. Its BLR size is also estimated to be $R_\mathrm{BLR}$ $\sim$ 10 to 20 ld \citep[e.g.][]{Du2014a, GRAVITY2023b}.

\subsection{IC 4329A}
IC 4329A is a type 1 Seyfert galaxy that is well-observed in X-ray \citep[e.g.,][]{Madejski1995, Delvaille1978, Piro1990, Nandra1994}. Several RM campaigns were implemented to measure the BLR size and BH mass of IC 4329A. For instance, \citet{Winge1996} presented the first RM data for IC 4329A, which was reanalysed by \citet{Peterson2004}. However, the poor quality of the light curves raised caution about their estimated BH masses and BLR sizes. \citet{Wandel1999} calculated a BLR size and BH mass of $R_\mathrm{BLR}$ = 1.4$^{+3.4}_{-2.9}$ ld and $M_\mathrm{BH} \sim 10^{7.4} \ M_\odot$ after re-analysing the spectrum taken by \citet{Winge1996}. \citet{Bentz2023} presented the latest RM campaign for IC 4329A, and together with BLR model fitting, they acquired a BLR size of $R_\mathrm{BLR}$ = 14.2$^{+7.2}_{-3.7}$ ld and a BH mass of $M_\mathrm{BH} = 4^{+10}_{-2} \times 10^{7.0} \ M_\odot$. \citet{GRAVITY2023b} also estimated the BLR size of IC 4329A via dust size measurements, yielding $R_\mathrm{BLR} \sim$ 17.4 ld.

\section{Observations and data reduction}
\label{sec:obs}

\subsection{Observations}
\begin{table*}
\caption{Observation logs of our target AGNs observed with GRAVITY. The table shows the dates, total on-source time of all the data used in the analysis after rejection via our fringe tracking ratio criterion, weather conditions (average seeing in arcseconds and coherence time in ms), and astrometric accuracy at the line peak (estimated from the photocentre fitting results; discussed more thoroughly in Sect. \ref{sec:photocentre}) of our observations.}           
\label{tab:obs_log}     
\centering                         
\begin{tabular}{|c|c|c|c|c|c|c}       
\hline        
\hline
Object & Dates & \makecell{Total on-source time \\ (min.)} & \makecell{Ave. Seeing \\ ($^{\prime\prime}$)} & \makecell{Ave. Coherence time \\ (ms)} &  \makecell{Astr. Accuracy \\ $\mu$as} \\
\hline
Mrk 509 & 25, 26 Jul. 2021 & 100 & 0.66 & 3.2 & 48.0\\
\hline
PDS 456 & \makecell{26, 27 Aug. 2018\\ 14, 15, 16, 18  Jul. 2019 \\ 12, 16 Aug. 2019 \\ 27 Jul. 2021} & 295 & 0.66 & 4.8 & 12.8 \\     
\hline
Mrk 1239 & \makecell{7, 30 Jan. 2021 \\ 1 Feb. 2021 \\ 1, 2, 31 Mar. 2021} & 366 & 0.72 & 6.8 & 32.7 \\   
\hline
IC 4329A & \makecell{1 Feb. 2021 \\ 1, 2, 31 Mar. 2021} & 210 & 0.76 & 5.2 & 47.4 \\    
\hline
\end{tabular}
\end{table*}

The GRAVITY observations were done over several nights between August 2017 and July 2021. We used the single-field on-axis mode with combined polarisation for all observations. Each observation sequence follows that described in \citet{GRAVITY2020a}. First, the telescopes pointed to the target and closed the adaptive optics (MACAO, \citealt{Arsenault2003}) loop. The light was then propagated to GRAVITY, where the fringe tracking (FT) and science channel (SC) fibres align on the target via internal beam tracking of GRAVITY. The exposures were acquired after the fringe tracker found the fringes and began tracking. The integration time for each exposure frame is DIT = 30 seconds, and the number of frames for each set of exposures is NDIT = 12. All observations were taken in MEDIUM resolution. The number of obtained exposures varied among objects. A calibrator star (usually an A- or B-type star) close to the target was also observed and used to calibrate the flux spectrum of the AGN. This ensures that atmospheric and vibrational effects, coherence loss, and birefringence are all accounted for. We adopted the same pipeline data reduction as \citet{GRAVITY2020a} to calculate the complex visibilities from the raw data, enabling us to extract quantities correlated to the physical properties of our targets. Table \ref{tab:obs_log} shows the date, exposure time and weather conditions (seeing and coherence time) during observations.

We use all available GRAVITY data for three of the four targets. However, we only use the observations of Mrk 509 from 2021. This is because earlier (2017 and 2018) observations of Mrk 509 show differential phase errors that are $\sim$50\% larger than those from 2021, which benefit from the factor two better throughput of the science channel spectrometer after the grating upgrade at the end of 2019 \citep{Yazici2021}. 

\subsection{Differential phase spectra}
To produce the differential phase curve on each baseline for all the targets, we first selected exposures with fringe tracking ratio (percentage of utilised time that fringe tracking was working) greater than or equal to 80\%, removing 16, 17, 8, and 33 exposures for Mrk 509, PDS 456, Mrk 1239, and IC 4329A, respectively. We followed the method from \cite{GRAVITY2020a} in removing instrumental features from the differential phase curves, estimating their uncertainties in each channel, and polynomial flattening. To check whether our uncertainties match the observed noise in the differential phase spectra and are a good representation of the actual dispersion of the differential phase spectra, we compare the average uncertainty in the whole spectra and the dispersion of the differential phase in regions beyond the wavelength range where the broad emission line is expected to be found. Our comparison suggests that only Mrk 1239 shows a significant discrepancy: the uncertainties in its differential phase spectra (which are $\sim$0.06-0.08$^\circ$ on average for all baselines) are lower compared to the dispersion of the differential phase values at around 2.18-2.20 $\mu$m and 2.22-2.24 $\mu$m (which are $\sim$0.08-0.14$^\circ$ on average). Hence, we adopted the dispersion of the differential phase values at these wavelength ranges as the representative uncertainty/error bar of all channels in each baseline. However, we emphasise that we also tried adopting the uncertainties in Mrk 1239's differential phase spectra as its representative uncertainty, and we conclude that this does not greatly affect our results.

The definition of differential phase in the partially resolved limit that is appropriate for our AGN observations \citep[e.g.][]{GRAVITY2018,GRAVITY2020a} is as follows:
\begin{equation}
    \Delta \phi_\lambda = -2\pi \frac{f_\lambda}{1 + f_\lambda} \vec{u} \cdot \vec{x}_{\mathrm{BLR}, \lambda},
    \label{eqn:visphi_cont}
\end{equation}
where $\Delta \phi_\lambda$ is the differential phase measured in a certain wavelength channel $\lambda$, $f_\lambda$ is the normalised flux in that channel, \vec{u} is the \textit{uv} coordinate of the baseline, and $\vec{x}_{\mathrm{BLR},\lambda}$ denotes the BLR photocentres. The differential phases are referenced to the continuum photocentre, which is placed at the origin. If the centre of the BLR coincides with the continuum photocentre, the differential phase signal reflects the kinematics of the BLR.  Meanwhile, if the BLR to continuum offset is much larger than the BLR size, the $\vec{x}_{\mathrm{BLR},\lambda}$ becomes a constant approximately over different channels. The differential phase signal will be proportional to $\frac{f_\lambda}{1+f_\lambda}$ and resemble the profile of the broad line (see more discussion in Section~\ref{sec:blr_offset_vs_lum}).  Hereafter, we call the phase signal generated by the global offset between the BLR and the continuum emission the ``continuum phase''.  Such a global offset is observed in IRAS 09149-6206 and NGC 3783 primarily due to the asymmetry of the hot dust emission \citep{GRAVITY2020a, GRAVITY2020b}. More details about the derivation of Eqn. \ref{eqn:visphi_cont} are shown in Appendix B of \cite{GRAVITY2020a}.

Fig. \ref{fig:visphi_4agns} shows the differential phase spectra of our 4 targets averaged for each baseline. Most of the baselines of the targets show signals in their differential phase curve that are either negative (Mrk 509 and PDS 456) or positive (Mrk 1239 and IC 4329A) and mostly coincide with the peak of their line profiles. As for IRAS 09149-6206 \citep{GRAVITY2020a}, this indicates that their differential phase spectra show signatures of a continuum phase. The signals are also relatively small with absolute values $\lesssim$ 0.6$^\circ$. We can, therefore, conclude that their BLRs are either intrinsically small or viewed at a very low inclination angle.

\begin{figure*}
    \centering
    \includegraphics[width=\textwidth]{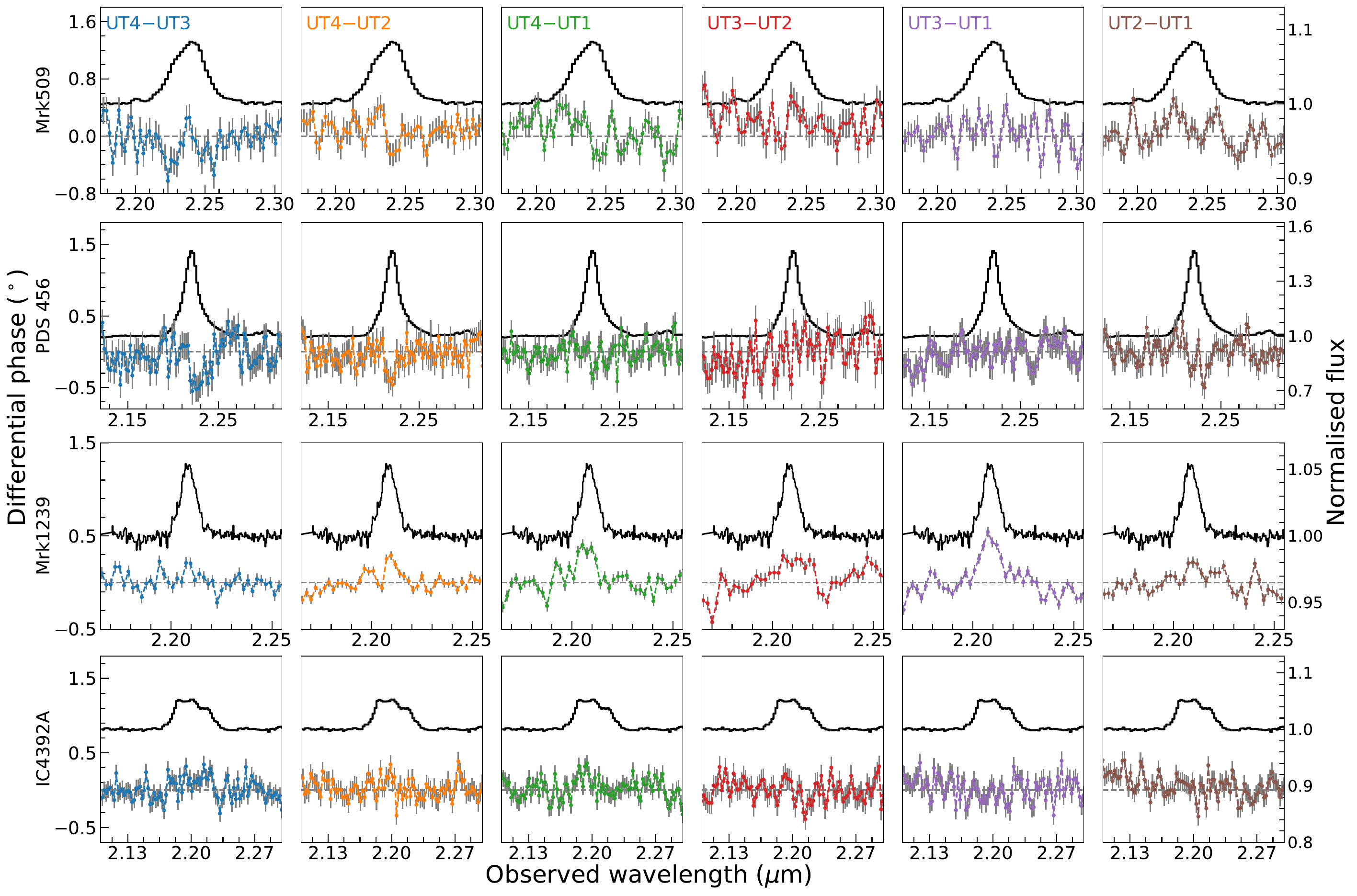}
    \caption{Averaged differential phase spectra (coloured, based on their respective baselines labelled in the top row) and the normalised Br$\gamma$ (for Mrk 509, Mrk 1239, and IC 4329A) and Pa$\alpha$ (for PDS 456) spectra (black histogram) of Mrk 509 (first row), PDS 456 (second row), Mrk 1239 (third row), and IC 4329A (fourth row). The curves show negative signals in Mrk 509 and PDS 456 with peaks ranging from $\sim$-0.4$^\circ$ to -0.5$^\circ$, and positive signals in Mrk 1239 and IC 4329A with peaks ranging from $\sim$0.5$^\circ$-0.6$^\circ$.}
    \label{fig:visphi_4agns}
\end{figure*}

\subsection{Normalised profiles of the broad Br$\gamma$ and Pa$\alpha$ emission lines}

Modelling the dynamics and deriving the velocity gradient of a target's BLR requires a well-measured nuclear emission line profile (Br$\gamma$ for Mrk 509, Mrk 1239, and IC 4329A, and Pa$\alpha$ for PDS 456). In general, the stacking and calibration of the line profiles are as follows: after fitting for and removing the Br$\gamma$ absorption line of the calibrator star, each spectrum was normalised by fitting a third-degree polynomial to the continuum before stacking. We then followed the method of extracting the final line profile from \citet{GRAVITY2020a} wherein we considered the statistical uncertainty (i.e. the rms from each individual spectrum's uncertainties) and the systematic uncertainty (i.e. mainly caused by variations in the calibrator data and sky absorption) of the spectra in producing the final flux error of the spectra. Lastly, the narrow components of the emission lines were also accounted for and removed following previous work \citep{GRAVITY`a}. Removing the narrow components from the flux spectra of our targets is important because the flux spectra, together with the differential phase spectra, are fitted together with our BLR model (to be discussed in Sect. \ref{sec:modelling}). Eqn. \ref{eqn:visphi_cont} assumes that all of the flux in the line profile is only partially resolved. Narrow line emission that enters the GRAVITY fibres occurs on large scales and is over-resolved and thus does not contribute to the differential phase. We note that the narrow component removal is only applied to the flux spectra and does not affect the interferometric data. We use the GRAVITY spectra of Mrk 509, PDS 456, and IC 4329A, while we use the spectrum from the Apache Point Observatory's (APO) TRIPLESPEC instrument of Mrk 1239. The APO spectrum of Mrk 1239 has a higher spectral resolution than that of GRAVITY, and we decided to use it to better remove the narrow component and characterise the line profile. Our comparison between the two Mrk 1239 spectra confirms that they are both consistent in terms of line shape and flux levels.

\begin{figure*}
    \centering
    \includegraphics[width=\textwidth]{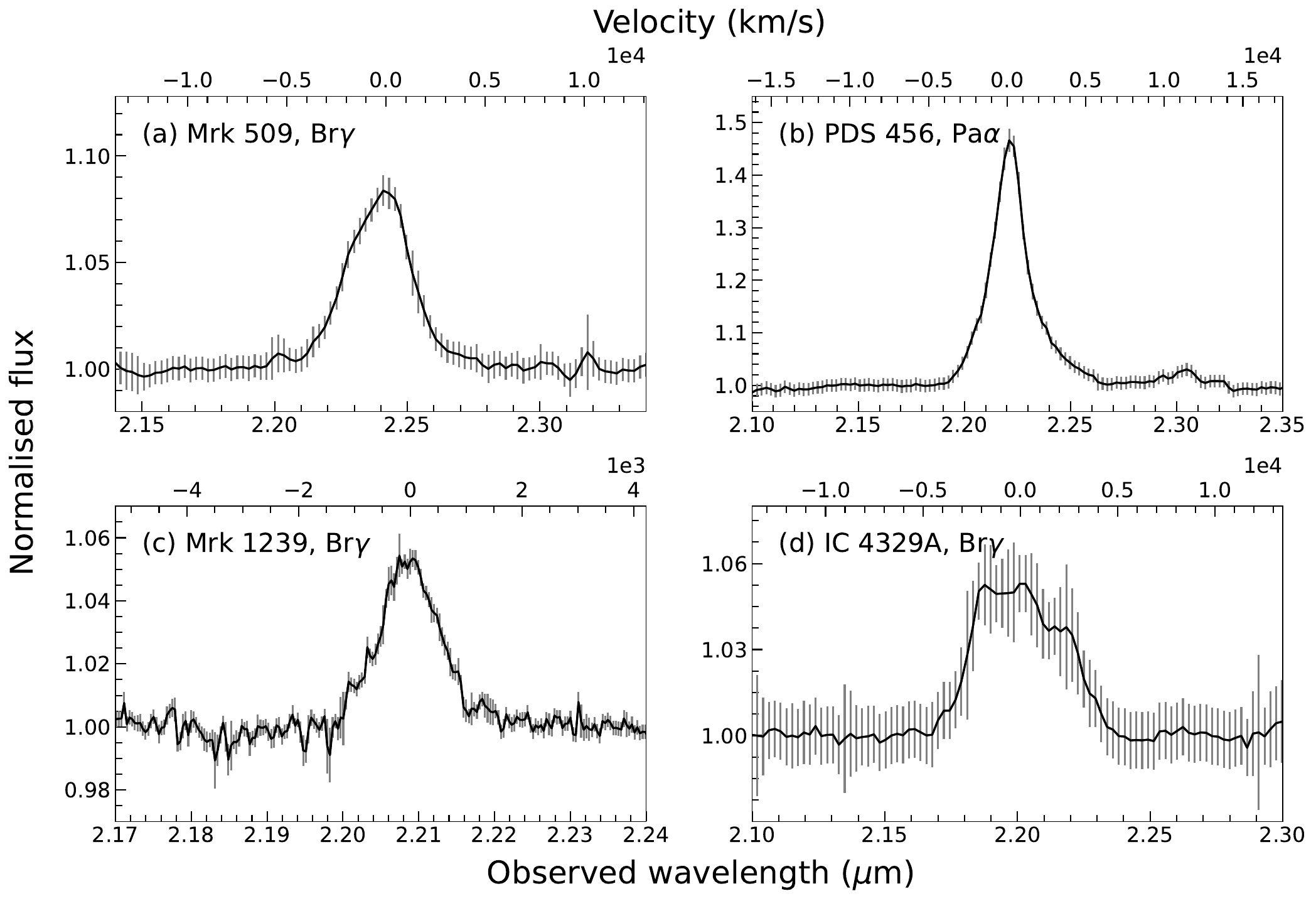}
    \caption{Average AGN flux spectra (black steps with grey error bars) of (a) Mrk 509 and (b) PDS 456 taken by GRAVITY, (c) Mrk 1239 taken by APO/TRIPLESPEC, and (d) IC 4329A taken by GRAVITY, all normalised to the continuum. The name of the emission lines for each target is shown in each panel.}
    \label{fig:all_spec}
\end{figure*}

\subsubsection{GRAVITY spectra: Mrk 509, PDS 456, and IC 4329A}
\label{sec:spec1}
The flux spectra (1.95-2.45 $\mu$m) of Mrk 509, PDS 456, and IC 4329A were taken with GRAVITY in medium resolution mode ($R$ = $\lambda$/$\Delta \lambda$ $\approx$ 500) with 90 independent spectral elements, which were extracted and resampled (and interpolated, if needed) into 210 channels. We extracted the final spectra of Mrk 509, PDS 456, and IC 4329A by weight-averaging 3, 10, and 3 GRAVITY spectra, respectively. Due to the low number of individual spectra to average for Mrk 509, we cannot simply estimate the flux errors in each channel based on multiple observations. Hence, we decided to multiply the flux errors by a factor of 3 to balance the weight of the flux spectrum and differential phase spectrum during BLR modelling. We also did the same for IC 4329A, but we decided to remove the factor of the flux error as it does not greatly affect the results of our analyses. To remove the Mrk 509's narrow Br$\gamma$ emission component, we used the optical spectrum of Mrk 509 taken by the Kitt Peak National Observatory (KPNO) 2.1m telescope \citep{Shang2005}. We utilised the optical blue spectrum (0.33-0.55 $\mu$m), adopting [O {\small{III}}]$\lambda$5008$\AA$ as our narrow line template. For PDS 456, we assume all of its light originates from the BLR. \citet{Simpson1999} showed very little [O {\small{III}}] emission in PDS 456's optical spectrum, which strongly suggests negligible narrow emission in its broad Pa$\alpha$ line. Hence, no narrow-line component removal was performed. For IC 4329A, no narrow line template was used to fit its narrow component. Although there is available VLT/SINFONI $K$-band (1.93 - 2.47 $\mu$m) spectrum for IC 4329A, it was taken last Jun. 2019, which is too distant in time for variability to be neglected. Therefore, we simply fit the Br$\gamma$ line profile with 6 Gaussian components and subtracted the component describing the narrow component at the central wavelength of Br$\gamma$. We note that for the flux spectrum of IC 4329A, any number of Gaussian components below 6 is insufficient to provide a good fit, while above 6 does not provide a better fit.

\subsubsection{APO and XSHOOTER spectra: Mrk 1239}
\label{sec:spec2}
We acquired 3 Mrk 1239 spectra (0.94-2.47 $\mu$m) with the TRIPLESPEC instrument at the Apache Point Observatory (APO) 3.5m telescope on two nights of 27th and 29th Dec., 2020. Since the GRAVITY interferometric observations were done $\sim$1-3 months after the APO spectroscopic observations, variability of spectral shape and intensity should be negligible as confirmed by \citet{Pan2021} after checking multi-epoch observations of Mrk 1239 in various wavelength ranges. The resolution of our Mrk 1239 flux spectrum is $R$ = 3181, almost 6 times that of GRAVITY's, making it a better choice for our analysis. We used a 1.1$"$ $\times$ 45$"$ slit for our APO observations. The spectra were reduced using the modified version of the \texttt{Spextool} package \citep{Cushing2004}. An A0V star was also used for telluric correction and flux calibration \citep{Vacca2003}.

To get a narrow line template suitable for removing the narrow component of Mrk 1239's Br$\gamma$ emission line, we observed Mrk 1239 with XSHOOTER in Dec. 2021 (PI: Shangguan, programme ID 108.23LY.001). The VIS (0.56-1.02 $\mu$m, $R$ = 18400, using slit dimensions of 0.7$''$ $\times$ 11$''$) spectrum of Mrk 1239 was reduced using the XSHOOTER pipeline version 3.5.3 running under the EsoReflex environment version 2.11.5. From this spectrum, we utilise [S{\tiny II}]$\lambda$6716$\AA$ as a template for the narrow component. 

We implement two methods to study the BLR of our targets: photocentre measurements (a model-independent method that reveals the BLR's possible velocity gradient and offset with respect to the hot dust) and BLR modelling (reveals BLR properties and central BH mass). More details about each method are discussed in Sects. \ref{sec:photocentre} and \ref{sec:modelling}.

\section{Measuring the BLR photocentres}
\label{sec:photocentre}

\begin{figure*}
    \centering
    \includegraphics[width=0.9\textwidth]{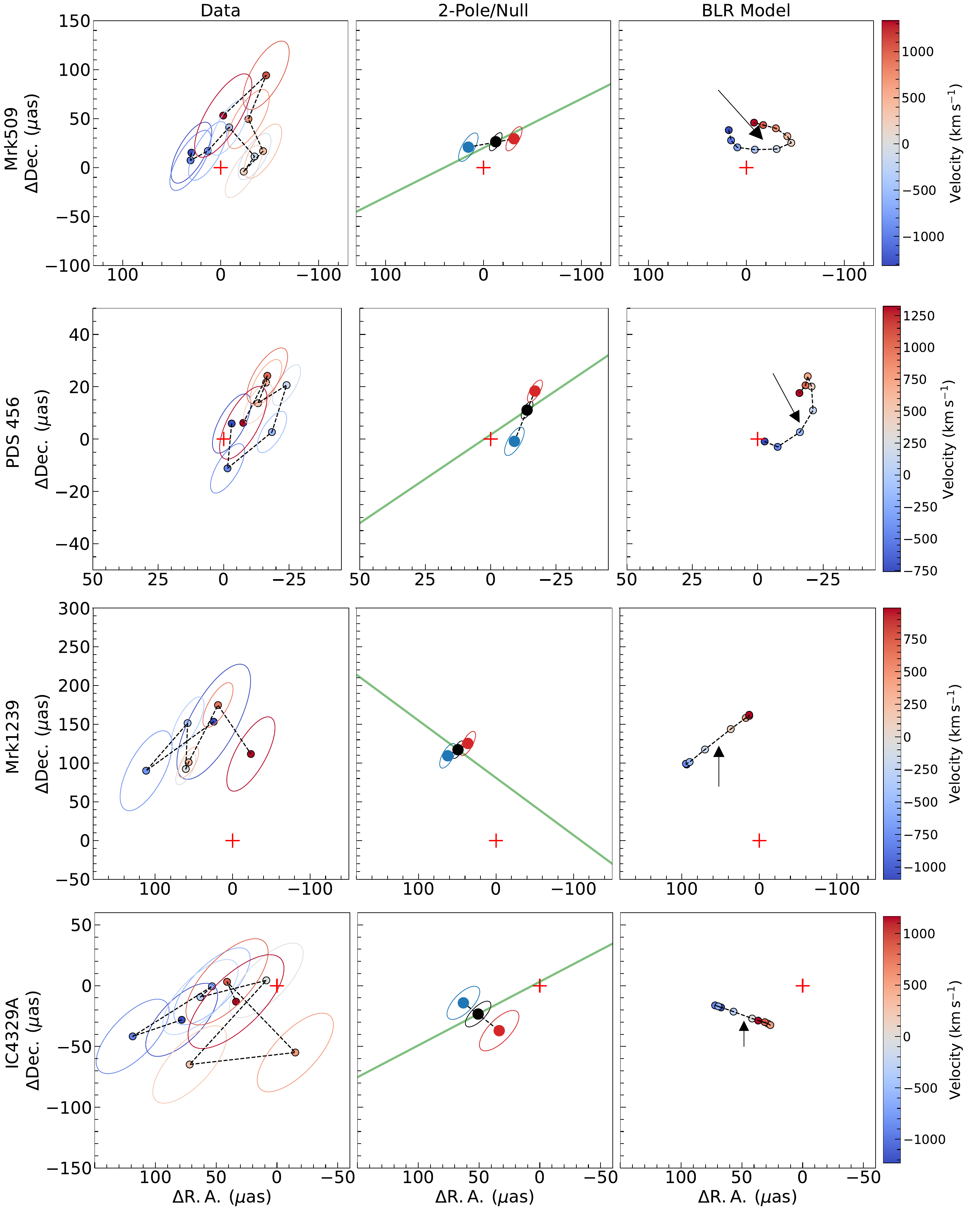}
    \caption{Best-fit BLR photocentres of our four targets. The columns from left to right show the photocentres from the data, the "2-pole" and "null" photocentre models (described in Sect.~\ref{sec:photocentre}), and the photocentres from the best-fit BLR model of each target. The red cross refers to the continuum photocentre. The direction of the radio jet is shown as a green solid line. The black arrow shows the position of the best-fit "null" model fitted from the best-fit BLR model. The ellipses around each centroid refer to 68\% (1$\sigma$) confidence intervals. The colours refer to the corresponding velocity of each spectral channel. The (significant) separation of the red- and blueshifted photocentres confirm that we have resolved the BLRs of our targets. For Mrk 1239 and IC 4329A, the perpendicular alignment between the radio jet and red-blue photo centres from the "2-pole" model fitting of the data and the straight alignments of their model photo centres indicate that their BLR is rotation-dominated. On the other hand, for Mrk 509 and PDS 456, the radio jet and red-blue photocentres are not closely perpendicular, and the model photocentres are curved, which suggests that their BLRs are radial motion-dominated.}
    \label{fig:photocentre_complete}
\end{figure*}

We start our analysis of the BLR structure by measuring the photocentre positions of the line emission. This model-independent method provides a direct representation of the differential phases in the various baselines. By measuring the BLR photocentres, we can confirm whether there is an overall velocity gradient in the BLR and investigate the offset between the BLR and hot dust photocentres, which previous works have also measured \citep{GRAVITY2020a, GRAVITY2021a}.

We first select the spectral channels where the Br$\gamma$/Pa$\alpha$ emission dominates. This criterion gives us 9, 8, 7, and 9 spectral channels to fit for Mrk 509, PDS 456, Mrk 1239, and IC 4329A, respectively. The photocentre displacement for each channel can then be calculated using Eqn. \ref{eqn:visphi_cont}. We also estimate the astrometric accuracy of each target (listed in Table \ref{tab:obs_log}) by averaging the errors of each photocentre position.

The left column of Fig.~\ref{fig:photocentre_complete} shows the BLR photocentres of the 4 targets as a function of wavelength. The photocentres show a velocity gradient in the BLR, indicated by the separation between the blueshifted and redshifted channels. To investigate the significance of this gradient, we fit all blueshifted and redshifted channels, assuming a single photocentre for each side. We call this our "2-pole" model. We also fit \textit{all} spectral channels into one photocentre, which we call our "null" model. The results of the 2-pole/null model fitting are shown in Fig. \ref{fig:photocentre_complete}, middle column. The resulting 2-pole model fittings reveal the separation between the red and blue poles (velocity gradient) of Mrk 509, PDS 456, Mrk 1239, and IC 4329A to be 48, 20, 30, and 37 $\mu$as, respectively. These correspond to 0.034, 0.064, 0.013, and 0.012 pc, respectively. An F-test was utilised to confirm the significance of the velocity gradient, with the null model as the null hypothesis and the 2-pole model as the alternative hypothesis.  Mrk 509, PDS 456, and IC 4329A reveal p-values that are $<<$ 0.05, indicating 6.7$\sigma$, 6.9$\sigma$, and 6.6$\sigma$ significance, respectively. However, the red-blue photocentre offset of Mrk 1239 has a p-value of 0.88, meaning we cannot reject the null hypothesis of a single photocentre describing the BLR. In this case, the confidence in the offset between the red and blue photocentres and the resulting position angle begs a more detailed model fitting. 

The BLR photocentres of our targets also reveal a systematic offset to the continuum photocentre (red cross in Fig. \ref{fig:photocentre_complete}). This offset, which we call the "BLR offset", is measured as the distance between the continuum photocentre and the BLR null photocentre, and is about 28, 17, 137, and 34 $\mu$as (0.020, 0.054, 0.057, and 0.011 pc) for Mrk 509, PDS 456, Mrk 1239, and IC 4329A, respectively. The offsets are mostly close to perpendicular to the BLRs' velocity gradient, which is similar to what was observed with IRAS 09149-6206 \citep{GRAVITY2020a}. To avoid the impact of possible data correlation between neighbouring channels during photocentre measurements, we re-binned the spectra by a factor of 2. The same method still yields a significance $>$4.5$\sigma$ (except Mrk 1239), regardless of the inclusion of the bluest/reddest channels that are furthest from the photocentres of the other channels.

On the right column of Fig. \ref{fig:photocentre_complete}, we also show the model photocentres, which are produced from the mock differential phase and flux spectra of the best-fit BLR model (See Sect. \ref{sec:discussion}). These model photocentres show the average photocentres of the BLR clouds of the best-fit BLR model in each wavelength channel and provide another way to see the cloud distribution, which can be linked to the differential phase data. These are discussed in more detail in Sect. \ref{sec:discussion}, where they are used to explain the observed differential phases of our targets.

Since we were able to constrain the BLR velocity gradient of most of our targets, we can compare the alignment of the red and blue photocentres with the jet alignment, as revealed by radio observations. The orientation of the radio jet is shown as the green line in the middle 2-pole/null model fitting results. The radio jet orientations are taken from \citet{Ulvestad1984} for Mrk 509, \citet{Yang2021} for PDS 456, \citet{Doi2015} and \citet{Orienti2010} for Mrk 1239, and \citet{Colbert1996} for IC 4329A. It is clear that the red-blue photocentre orientations of Mrk 1239's and IC 4329A's BLRs are perpendicular to their radio jets, indicating that their BLRs are rotating, similar to what is found for 3C 273 \citep{GRAVITY2018}. However, Mrk 509 and PDS 456 reveal that the red-blue photocentre orientations of their BLRs are closely aligned to their radio jets.

To conclude our photocentre fitting analysis, we find significant ($> 5\sigma$) velocity gradients between the blueshifted and redshifted channels of all our targets except Mrk 1239. These velocity gradients are perpendicular to the BLR offsets. Our comparison between the red and blue photocentres and the orientation of the radio jets shows that the BLRs of Mrk 1239 and IC 4329A are rotating. However, Mrk 509 and PDS 456 show otherwise. Photocentre fitting cannot explain this phenomenon, let alone fully constrain the physical BLR size, as the photocentre offsets average over the emission of the channel bandpass and will therefore underestimate the BLR size. Therefore, a flexible model fit to the full differential phase spectra must be utilised to investigate our targets' BLR structure and kinematics. 

\section{Description of the BLR model}
\label{sec:modelling}

We follow the BLR model described in \citet{Pancoast2014} (henceforth called the \textit{Pancoast model}). This model defines the BLR as a collection of non-interacting clouds encircling the central SMBH. This is a convenient way to model the kinematics and structure of the BLR, but we do not mean to imply that the BLR specifically comprises discrete clouds. The physical structure of the BLR is described by several parameters that dictate the clouds' position and motion. Table 2 of \citet{GRAVITY2020a} summarises the parameters included in the model and their possible ranges of values. We use the same priors as \citet{GRAVITY2020a}, with a few exceptions, which are discussed later. Hence, we refer the readers to \citet{GRAVITY2020a} for the full list of parameters in the model.

In the model, $R_\mathrm{BLR}$ is the average (emissivity-weighted) BLR radius, and $\beta$ controls the radial cloud distribution, which can be Gaussian (0 $<$ $\beta$ $<$ 1), exponential ($\beta$ = 1), or heavy-tailed and steep inner profile (1 $<$ $\beta$ $<$ 2). The fractional inner radius is defined as $R_{\rm min}/R_{\rm BLR}$, where $R_\mathrm{min}$ is the minimum or inner radius of the BLR. The BLR can be inclined relative to the plane of the sky by an inclination angle ($i$) defined such that face-on is $i=0^\circ$ and edge-on is $i=90^\circ$. The half-opening angle of the disc, $\theta_0$, describes the thickness of the BLR: $\theta_0 = 90^\circ$ produces a spherical BLR, while $\theta_0 = 0^\circ$ produces a thin disc. Finally, the BLR is rotated within the plane of the sky by the position angle, $PA$. The parameter $\gamma$ controls the angular concentration of BLR clouds relative to $\theta_0$. When set to 1, it pertains to the uniform case with the clouds equally distributed as a function of angular height. Increasing $\gamma > 1$ redistributes the BLR clouds more and more to the outer faces of the disc, with $\gamma = 5$ as the maximum possible value.

Asymmetries can also be introduced in the BLR model through two effects. We can introduce anisotropic emission from each individual BLR cloud via the parameter $\kappa$. We assign a weight "$w$" to each cloud that defines how much BLR emission is directed into the LOS:
\begin{equation}
    \label{weight}
    w = 0.5 + \kappa \cos \phi
\end{equation}
where $\kappa$ ranges from -0.5 to +0.5. When $\kappa > 0$, it results in preferential emission from the BLR's near side (side closer to the observer), while $\kappa < 0$ results in preferential emission from the BLR's far side. The parameter $\phi$ describes the angle between the LOS of the observer and the LOS of the BLR cloud to the central ionising source. Anisotropic emission from the near side could be caused by BLR clouds situated closer to the observer obstructing the emission from the gas farther away. On the other hand, preferential emission from the far side of the BLR could be caused by self-shielding within individual BLR clouds, resulting in emission only at the back towards the central continuum source.

Finally, the transparency of the midplane can be modelled with the parameter $\xi$, which ranges from 0 to 1. When $\xi$ = 1, the clouds are evenly distributed on both sides of the equatorial plane, while $\xi$ = 0 means that the emission from the clouds behind the equatorial plane is obscured. The physical cause of possible BLR mid-plane opacity is not well-understood \citep{GRAVITY2020a}, and it should be interpreted cautiously. 

Two parameters control the effect of radial motion in the BLR clouds: $f_\mathrm{ellip}$ and $f_\mathrm{flow}$. $f_\mathrm{ellip}$ is the fraction of BLR clouds that have bound elliptical orbits. The rest (1-$f_\mathrm{ellip}$) have much more elongated orbits dominated by radial motion. $f_\mathrm{flow}$ is a flag that controls whether these elongated orbits are inflowing (0 $<$ $f_\mathrm{flow}$ $<$ 0.5) or outflowing (0.5 $<$ $f_\mathrm{flow}$ $<$ 1). The parameter $f_\mathrm{flow}$ acts as a "binary flag" and is thus not a continuous variable. We also fit the central wavelength of the line, the line peak flux, and the BLR centre. However, we do not focus our discussion on these parameters.

We consider two variations of the Pancoast model in our work. One is termed the \textit{circular} model. This model sets $\kappa$ = 0 and $\xi$ = 1 to ensure that all clouds emit isotropically and are uniformly distributed above and below the BLR mid-plane. No angular asymmetry is considered in this model; that is, $\gamma$ = 1. All BLR clouds are then subject to circular Keplerian rotation (i.e., $f_\mathrm{ellip}$ = 1). The second model is the \textit{elliptical/radial} model. In this model, angular asymmetry ($\gamma$) and all parameters corresponding to the asymmetrical properties of the BLR ($\kappa$, $\xi$) are fitted. Inflowing/outflowing clouds ($f_\mathrm{ellip}$ and $f_\mathrm{flow}$) are fitted as well (together with the Keplerian clouds). We fit the circular and elliptical/radial models to our 4 targets, but we only show the best-fit results for each target. We discuss the reason for our choice of BLR model during the fitting of each target in Sect. \ref{sec:discussion}. 

For the exceptions compared to \citet{GRAVITY2020a}, we fix the angular and radial standard deviations of circular and radial orbit distributions and the standard deviation of turbulent velocities to zero because they do not affect the model fitting. We limit the parameter space of the inclination angle to be Uniform($\cos$ $i$(0, $\pi$/4)). We tested the fitting for all targets with a much larger range, Uniform($\cos$ $i$(0, $\pi$/2)), and found that all targets except Mrk 509 and IC 4329A prefer a smaller inclination. Hence, we decided to use Uniform($\cos$ $i$(0, $\pi$/2)) for the parameter space of the inclination angle of Mrk 509 and IC 4329A, while we adopt the smaller parameter space, Uniform($\cos$ $i$(0, $\pi$/4)), to the rest of the targets to avoid multiple peaks in the posterior distribution. More discussion about the inclination angle of Mrk 509 and IC 4329A is presented later (in Sect. \ref{sec:disc_mrk509} and \ref{disc_ic4329a}, respectively).

We fit the models to the total flux line profile and differential phase curves from all baselines simultaneously. The chosen wavelength channels for fitting each target cover the wavelength range where Br$\gamma$/Pa$\alpha$ emission flux is relevant. All models utilise 2 $\times$ 10$^5$ clouds, which are randomly given parameter values based on their model distributions.

The interferometric data and prior information about the source and model are used to infer the best-fit model parameters via Bayes' theorem. We use the Python package, \texttt{dynesty} \citep{Speagle2020}, to implement posterior distribution sampling (random walk) of 2000 live points for each BLR fitting. \texttt{dynesty} allows the nested sampling algorithm to be utilised, which is powerful in estimating Bayesian evidence and dealing with complex models with presumably multimodel posteriors. The Bayesian evidence or marginal likelihood, $Z$, measures how well the probability distributions are constrained. The larger the $Z$, the stronger the constraint to the probability distributions. The Bayes factor, or the ratio of the Bayesian evidences, is then calculated to identify which model is more apt to fit the data. 

We prefer to fit two models (inflowing and outflowing elliptical/radial model) separately and compare their Bayes factor, or the ratio of the outflowing model evidence to the inflowing model evidence, due to the nature of $f_\mathrm{flow}$ parameter as a binary flag. We do this by fitting an elliptical/radial model with $f_\mathrm{flow}$ restricted to a value between 0-0.5 (for inflowing clouds, representing an inflow elliptical/radial model) or 0.5-1.0 (for outflowing clouds, representing an outflowing elliptical/radial model). If the Bayes factor $>$ 1, the source prefers outflowing radial motions; otherwise, it prefers inflowing radial motions. For all of our fittings, we use the nesting sampling algorithm (\texttt{NestedSampler}) as the sampler and random walk (\texttt{rwalk}) as the sampling method.

In this work, we report the best-fit values and their 68\% (1$\sigma$) credible intervals of each parameter. We also show the model-inferred photocentre fitting results in Fig. \ref{fig:photocentre_complete}, right column, where we simulate the differential phase and flux spectra from the best-fit BLR models of our targets. We observe that our model-inferred photocentre fitting results are consistent with what we get from fitting the photocentre from our data.

\section{Importance of differential phase data in BLR fitting}
\label{sec:visphi_importance}

\subsection{Resulting posterior distribution from BLR fitting with and without phase data}
\label{sec:expected_posterior}

\begin{figure}
    \includegraphics[width=0.45\textwidth]{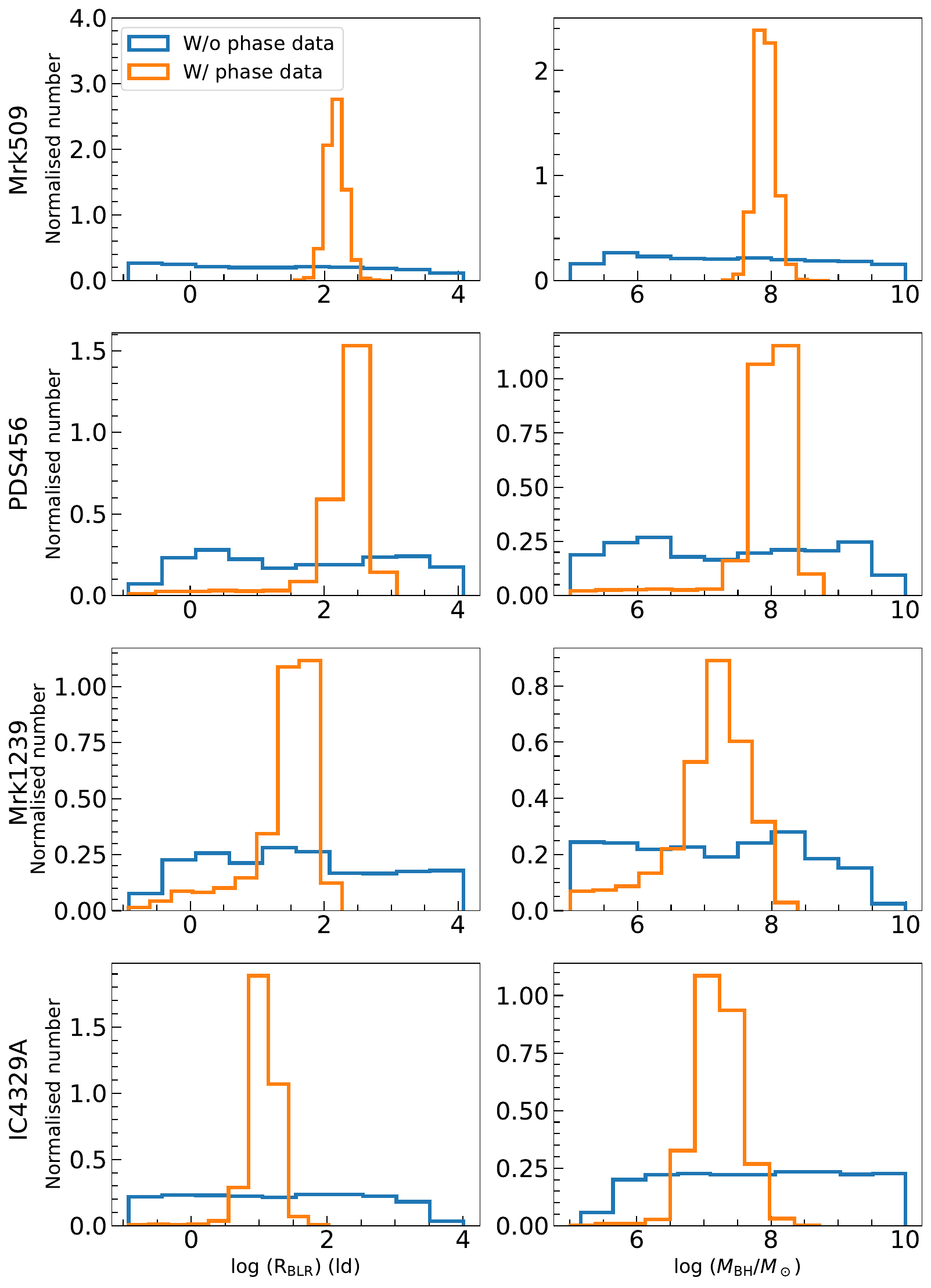}
    \caption{Posterior distributions of BLR radius (left column) and BH mass (right column) resulting from BLR fitting without differential phase data (blue histogram) and with differential phase data (orange histogram). Each row corresponds to each target; from top to bottom: Mrk 509, PDS 456, Mrk 1239, and IC 4329A. The histograms are normalised such that they integrate into 1. It is clear that the spatial information from the differential phase significantly constrains both the BLR radius and BH mass.} 
    \label{fig:with_vs_without_phase}
\end{figure}

\begin{figure*}
    \includegraphics[width=\textwidth]{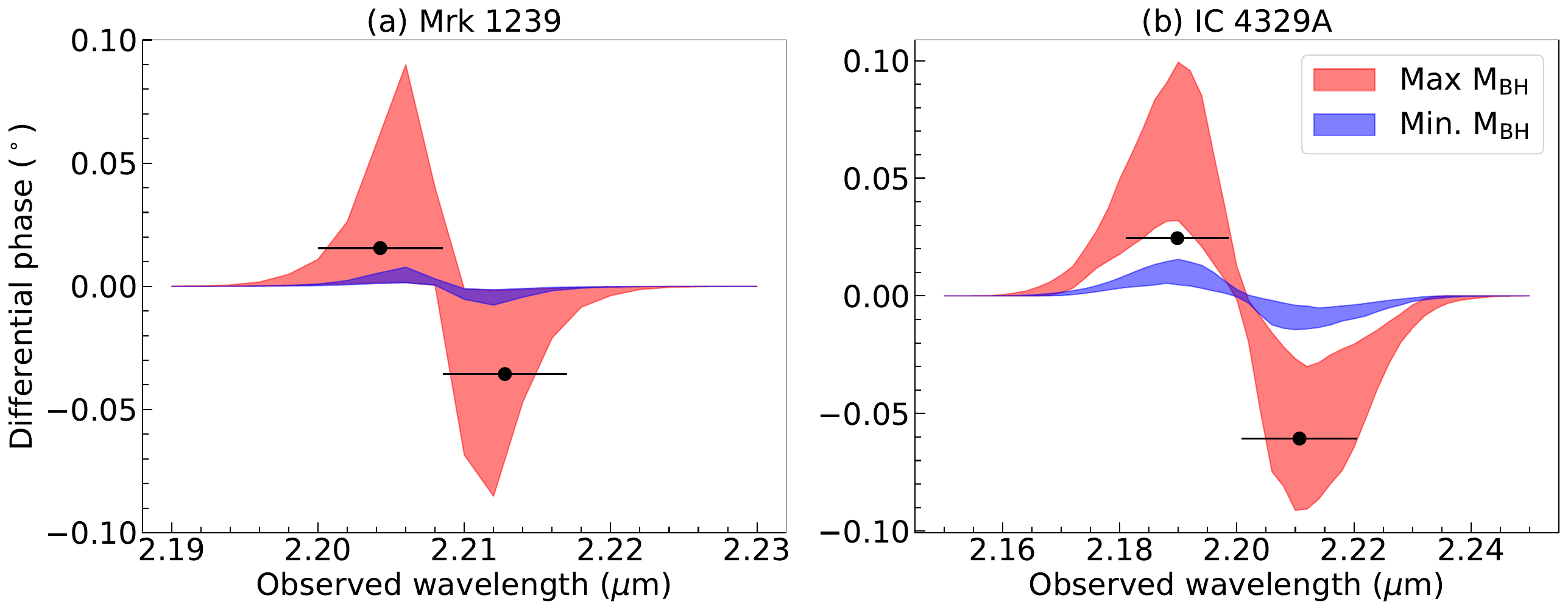}
    \caption{Expected differential phase signals for Mrk 1239 (left panel) and IC 4329A (right panel) when their line profiles were fitted with our circular model assuming their maximum (red shaded region) and minimum (blue shaded region) BH masses were fixed during fitting. The assumed BH masses for Mrk 1239 and IC 4329A are taken from previous literature as discussed in the text. The shaded regions refer to 1$\sigma$ uncertainty. The chosen baselines to average are the same as the ones used to create Fig. \ref{fig:all_ave_visphi}: UT4-UT3, UT4-UT2, and UT4-UT1 for Mrk 1239; UT4-UT2, UT4-UT1, and UT3-UT2 for IC 4329A. The black points refer to the maximum and minimum observed phase signals at the line region of Mrk 1239 and IC 4329A after averaging the differential phase spectra from the same chosen baselines. The horizontal lines show the wavelength range of the maximum and minimum observed phase signals.}
    \label{fig:expected_phase_signals}
\end{figure*}

As discussed in the previous section, the differential phase and flux spectra are our key inputs for BLR fitting. The differential phase measures the broad emission line's photocentre shift at different wavelength channels with respect to the continuum, providing spatial information. On the other hand, the flux spectrum gives information on the distribution of gas velocities in the BLR (see \citealt{Raimundo2019}, \citealt{Raimundo2020}, and references therein), as well as the inclination angle of the BLR, which is degenerate with the observed BH mass \citep{Rakshit2015, GRAVITY2018}. The work of \citet{Pancoast2014}, from which our model is derived, performed fitting of broad emission lines (usually H$\beta$). In this section, we investigate the importance of adding the differential phase spectrum in fitting the broad emission line with our BLR model.

First, we fit our four targets with and without their differential phase spectra, assuming the same priors as the fitting with the differential phase spectra, and then compare their posterior distributions. We focus on the posterior distributions of the BLR radius, log $R_\mathrm{BLR}$ [ld], and BH mass, log $M_\mathrm{BH}$ [$M_\odot]$ since these are the parameters that cannot be constrained with only a line profile. Fig. \ref{fig:with_vs_without_phase} shows the comparison between the posterior distributions of our BLR fitting with (orange line) and without (blue line) differential phase spectra for our 4 targets. It is clear that without the phase data, our BLR fitting fails to constrain these parameters. Our simple comparison shows that the phase data is crucial for unveiling more accurate estimates of our targets' BLR size and BH mass and more precise pictures of their BLR geometry and kinematics. This still holds true even for cases when there is no differential phase signal detected.

\subsection{Expected differential phase signals of Mrk 1239 and IC 4329A}
\label{sec:expected_visphi}

In the previous subsection, we highlight the importance of differential phase data in our BLR model fitting even if there is no phase signal detected. We investigated further by estimating the expected phase signals for Mrk 1239 and IC 4329A if only their flux profile were available for fitting. We focus on these two objects because, among our 4 targets, they have very weak signals that are still below their noise level ($\sim$0.05$^\circ$ and $\sim$0.10$^\circ$ for Mrk 1239 and IC 4329A, respectively). To constrain the possible phase signals of Mrk 1239 and IC 4329A without the phase data, we ran our BLR fitting for these two objects without their differential phase spectra while fixing their BH masses into their possible maximum and minimum values from previous works. For Mrk 1239, the maximum BH mass is log $M_\mathrm{BH}$ [$M_\odot]$ = 7 from \citet{Pan2021}, while the minimum BH mass is log $M_\mathrm{BH}$ [$M_\odot]$ = 5.9 from \citet{Kaspi2005} and \citet{Greene2005}. For IC 4329A, the maximum BH mass is log $M_\mathrm{BH}$ [$M_\odot]$ = 7.6 from \citet{Bentz2023}, while the minimum BH mass is log $M_\mathrm{BH}$ [$M_\odot]$ = 6.8 from \citet{Kaspi2000}. The expected averaged phase signals are shown in Fig. \ref{fig:expected_phase_signals} as a 1$\sigma$ shaded region. The chosen baselines to average are UT4-UT3, UT4-UT2, and UT4-UT1 for Mrk 1239, and UT4-UT2, UT4-UT1, and UT3-UT2 for IC 4329A. To create Fig. \ref{fig:expected_phase_signals}, the $PA$ of each object were also fixed to their best-fit values (see best-fit results in Table \ref{tab:circular}). Since there is a degeneracy between the BH mass, BLR size, and inclination angle \citep{Rakshit2015, GRAVITY2018}, the expected phase signal will then reflect the possible BLR sizes and inclination angles for these targets; that is, the larger the expected phase signal is, the bigger the inferred BLR size and inclination angle are.

Fig. \ref{fig:expected_phase_signals} shows the range of differential phase signals that we are expecting based on what was previously known. The expected phase signals for the two objects span a large range when the maximum BH mass is assumed, indicating the wide range of possible BLR radius and inclination angles. The resulting absolute phase signal goes as high as $\sim$0.09$^\circ$ for Mrk 1239 and $\sim$0.10$^\circ$ for IC 4329A. The opposite is true when the minimum BH mass is assumed; the resulting absolute phase signals are very small ($\sim$0.005$^\circ$ for Mrk 1239, and $\sim$0.02$^\circ$ for IC 4329A) and cover a very narrow area in the differential phase parameter space. The areas covering the expected phase signals for Mrk 1239 and IC 4329A when the maximum BH masses are assumed are much larger than the typical signals shown in the differential phase spectra of these objects (see Fig. \ref{fig:visphi_4agns}). This emphasises the importance of phase data in our fitting; without the phase data, the constraining power of our fitting is lessened, and the resulting posterior distributions of their BLR radius and inclination angle span a much wider range. The rather weak signals of Mrk 1239 and IC 4329A have a discriminating power, allowing us to limit the possible values of their BLR radii and inclination angles. Another way to look at this is that the small phase signals of Mrk 1239 and IC 4329A rule out higher values of inclination angle because these would lead to stronger phase signals.

The expected phase signals of Mrk 1239 and IC 4329A could also be used to infer the possible BH masses of these targets by comparing them with their observed phase signals. In Fig. \ref{fig:expected_phase_signals}, we plot the average differential phase on the line region channels located at the left and right of the line centre. These represent the average observed phase signals of Mrk 1239 and IC 4329A. For Mrk 1239, its average observed phase signal runs from -0.04$^\circ$ to +0.02$^\circ$. Based on the expected phase signals in Fig. \ref{fig:expected_phase_signals}, this indicates that Mrk 1239 may not have as low of a BH mass as \citet{Kaspi2005} measured, and instead, it may have a large BH mass close to what \citet{Pan2021} measured. The average observed phase signal of IC 4329A (from -0.06$^\circ$ to +0.02$^\circ$) also suggests that IC 4329A may not possess such a low BH mass similar to what \citet{Kaspi2000} measured. This indicates that IC 4329A may have a large BH mass similar to the value \citet{Bentz2023} measured. The BLR fitting results discussed next section will confirm if these points are true or not.

\section{BLR modelling results}
\label{sec:discussion}

\begin{figure*}
    \centering
    \includegraphics[width=\textwidth]{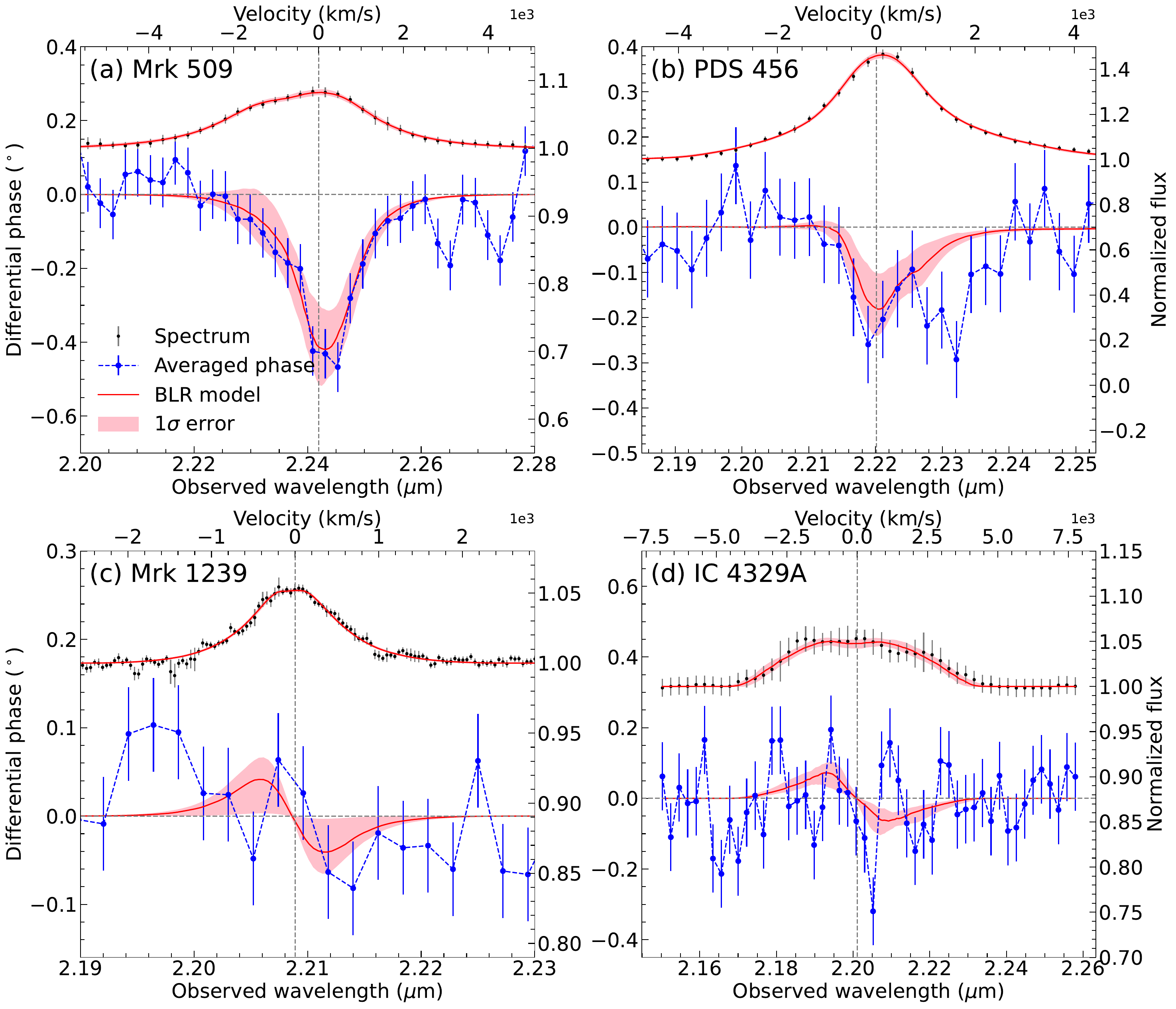}
    \caption{Summary of BLR fitting results for (a) Mrk 509, (b) PDS 456, (c) Mrk 1239, and (d) IC 4329A. The observed averaged differential phase spectra with the continuum phase signal removed are shown in blue, while the median differential phase spectra from the best-fit BLR model are shown in red (see text for the details of averaging the differential phase spectra for each target). Above each panel, the observed line profile (black) is shown together with the median best-fit model line profile (red). The central wavelengths of the line profiles are shown by the grey vertical dashed lines. The red-shaded region shows the 16-to-84 percentiles of the averaged differential phase spectra and flux spectra from the best-fit model, which was created by getting randomly selected samples of each parameter in 100 different instances and then recreating the averaged differential phase spectra. Mrk 509 and PDS 456 clearly show an asymmetrical signal typical for the Pancoast model, while Mrk 1239 and IC 4329A show the S-shape signal typical for the Keplerian model.}
    \label{fig:all_ave_visphi}
\end{figure*}

Fig. \ref{fig:all_ave_visphi} summarises our BLR modelling results. We compare the differential phase spectra from our data with the ones predicted by the model, using the weighted average of the differential phase spectra from the longest baselines and/or the baselines that show the strongest or most prominent BLR phase signal after continuum phase removal (note that all the baselines were used for the actual BLR fitting; we only do this step for clearly visualising the averaged phase of each target). We show the 16-to-84 percentile range (1$\sigma$ confidence interval) of the model phase and flux spectra as a red-shaded region by randomly sampling the posterior distribution of our fitting results 100 times.

The differential phase signal will vary over different baselines due to their different orientation with respect to the observed target. Therefore, for visualisation purposes, we have the freedom to choose which baselines we should average to compare the data and model fitting results. For Mrk 1239, the averaged differential phase spectrum is derived from averaging the differential phase spectra from three baselines: UT4-UT2, UT4-UT1, and UT4-UT3. For IC 4329A, the chosen baselines are similar to that of Mrk 1239, except that UT3-UT1 is chosen instead of UT4-UT3. For Mrk 509 and PDS 456, we averaged the spectra from all baselines. The observed differential phase includes both the continuum phase and the BLR differential phase. We are only interested in the latter. Hence, we subtract the former based on our model fitting. Mrk 509 and PDS 456 show a strong asymmetric signal in their averaged BLR differential phase. Mrk 509 also shows a slight asymmetry in its flux spectrum. On the other hand, Mrk 1239 and IC 4329A reveal weak BLR signals, which are below their respective noise. Considering this, we decided to fit the elliptical/radial model for Mrk 509 and PDS 456 data, while we fit the circular model for Mrk 1239 and IC 4329A data. The low signal-to-noise ratio (S/N) data of Mrk 1239 and IC 4329A cannot be well-constrained by the elliptical/radial model, and even if we fit them with the elliptical/radial model, the resulting BH mass and BLR size are not much different given our uncertainties, similar to what is found for IRAS 09149-6206 in \citet{GRAVITY2020a}.

Tables \ref{tab:pancoast} and \ref{tab:circular} show the best-fitting parameters of the elliptical/radial and circular models for our 4 targets. We discuss each target's BLR modelling results (i.e., averaged differential phase spectra, best-fit parameter values, and their interpretation) in the following subsections.

\begin{table}[htp]
\centering
\caption{Inferred maximum a posteriori value and central 68\% credible interval for the modelling of the spectrum and differential phase of Mrk 509 and PDS 456 with the elliptical and radial model.}\label{tab:pancoast}

\begin{tabular}{|c|c|c|c|c}       
\hline   
\hline
Parameter & Mrk 509 & PDS 456 \\
\hline
log $R_\mathrm{BLR}$ [ld]           & $2.29^{+0.01}_{-0.26}$ & $2.49^{+0.08}_{-0.38}$ \\
log R$_{\rm BLR,\ min}$ [ld]        & $0.93^{+0.40}_{-0.39}$ & $1.16^{+0.42}_{-0.30}$ \\
$\beta$                             & $1.07^{+0.15}_{-0.14}$ & $1.83^{+0.06}_{-0.20}$ \\
$i [^\circ]$                        & $69^{+6}_{-12}$ & $13^{+9}_{-2}$ \\
$PA [^\circ]$                       & $185^{+25}_{-7}$ & $265^{+2}_{-214}$ \\
$\theta_0 [^\circ]$                 & $64^{+11}_{-9}$ & $42^{+14}_{-6}$ \\
log $M_\mathrm{BH}$ [$M_\odot]$     & $8.00^{+0.06}_{-0.23}$ & $8.23^{+0.01}_{-0.49}$ \\
$\gamma$                            & $4.1^{+0.3}_{-2.4}$ & $1.55^{+1.37}_{-0.18}$ \\
$\kappa$                            & $-0.18^{+0.09}_{-0.11}$ & $-0.44^{+0.07}_{-0.04}$ \\
$\xi$                               & $0.21^{+0.10}_{-0.17}$ & $0.75^{+0.11}_{-0.10}$ \\
$f_\mathrm{ellip}$                  & $0.30^{+0.13}_{-0.10}$ & $0.52^{+0.04}_{-0.25}$ \\
$x_0$ [$\mu$as]                     & $-3.6^{+26.3}_{-13.4}$ & $-7.3^{+3.4}_{-2.6}$ \\
$y_0$ [$\mu$as]                     & $166.9^{+4.8}_{-65.7}$ & $13.6^{+0.7}_{-5.8}$ \\
\hline

\end{tabular}
\begin{flushleft}
\textbf{Note:} 
log R$_{\rm BLR, \ min}$ was calculated using the formula $F$ = $R_{\rm BLR, \ min}/R_{\rm BLR}$, where both $F$ and $R_\mathrm{BLR}$ are both fitted parameters.
\end{flushleft}

\bigskip

\small
\centering
\caption{Similar to Table \ref{tab:pancoast}, but for Mrk 1239 and IC 4329A with the circular model.}
\label{tab:circular}

\begin{tabular}{|c|c|c|c|c}   
\hline
\hline
Parameter & Mrk 1239 & IC 4329A \\
\hline
log $R_\mathrm{BLR}$ [ld]         & $1.77^{+0.03}_{-0.74}$ & $1.13^{+0.10}_{-0.23}$ \\
log R$_{\rm BLR,\ min}$ [ld]      & $0.64^{+0.27}_{-0.28}$ & $0.64^{+0.27}_{-0.28}$ \\
$\beta$                           & $1.21^{+0.29}_{-0.31}$ & $1.81^{+0.09}_{-0.92}$ \\
$i [^\circ]$                      & $11^{+6}_{-3}$ & $54^{+22}_{-20}$ \\
$PA [^\circ]$                     & $197^{+19}_{-49}$ & $155^{+8}_{-53}$ \\
$\theta_0 [^\circ]$               & $42^{+18}_{-15}$ & $54^{+26}_{-29}$ \\
log $M_\mathrm{BH}$ [$M_\odot]$   & $7.47^{+0.15}_{-0.92}$ & $7.15^{+0.38}_{-0.26}$ \\
$x_0$ [$\mu$as]                   & $55.1^{+4.2}_{-10.0}$ & $47.0^{+10.7}_{-5.8}$ \\
$y_0$ [$\mu$as]                   & $130.0^{+11.3}_{-11.6}$ & $-25.9^{+6.0}_{-10.0}$ \\
\hline
\end{tabular}
\end{table}

\subsection{Mrk 509}
\label{sec:disc_mrk509}

Mrk 509 is best fit with the elliptical/radial model, which better reproduces its asymmetric flux spectrum. The Bayesian evidence from the elliptical/radial Pancoast model, which is $\sim100 \times$ greater than that from the circular Pancoast model, strongly indicates the need for including both asymmetric emission and radial motion. Fig. \ref{fig:corner_plot_mrk509} in the Appendix shows the corner plot of the resulting posterior distributions of all fitted parameters. The inferred inclination angle is $\sim$ 69$^\circ$. Although it is quite high for a face-on target, we note that with the presence of radially moving clouds, the definition of the inclination angle is not as straightforward as that of a BLR possessing only rotation-dominated clouds.

\begin{figure*}
    \centering
    \includegraphics[width=\textwidth]{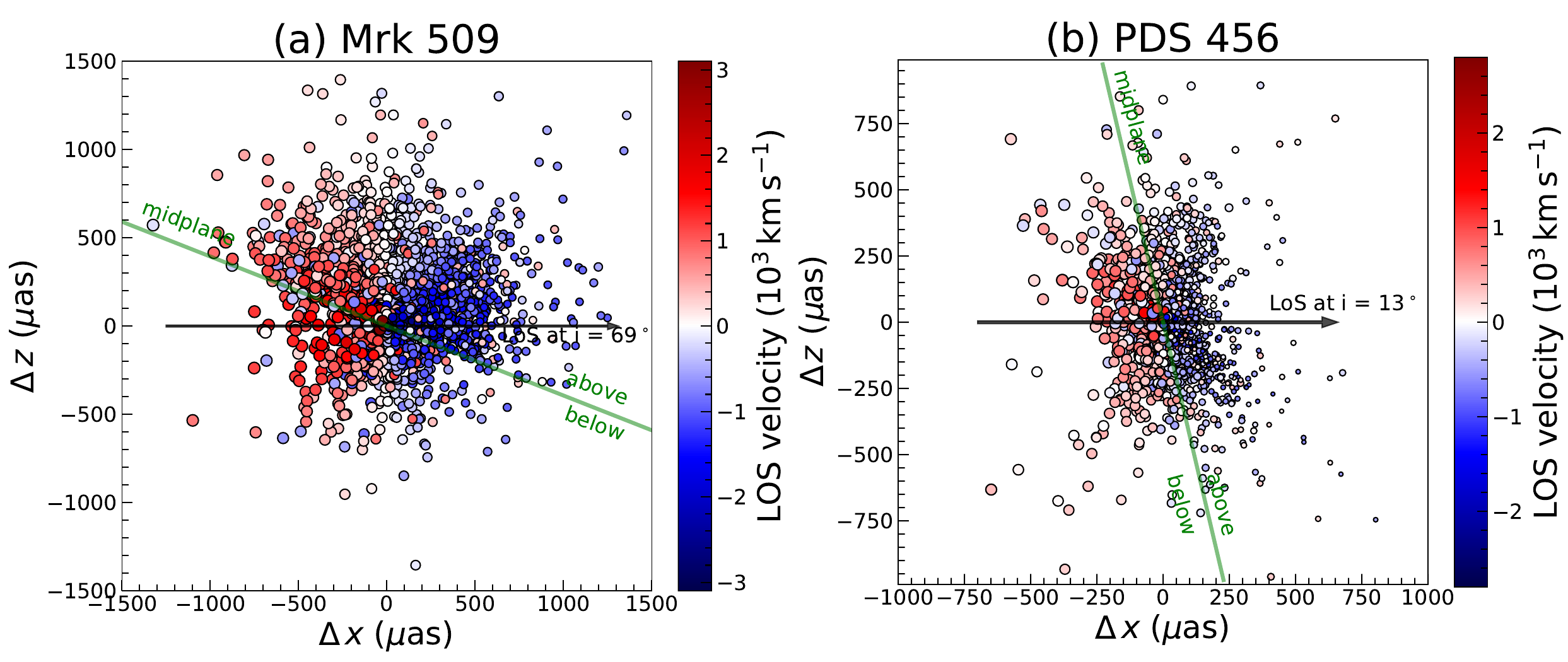}
    \caption{Edge-on views of the best-fit outflow model for (a) Mrk 509 and (b) PDS 456. For clarity, the PA for both configurations is adjusted to 180$^\circ$, and the BLR centre is positioned to the origin.
    The LOS at the best-fit inclination angle is shown as a black arrow. The midplane, represented as a green line, is defined as the line passing through the origin and is perpendicular to the LOS at $i$ = 0$^\circ$. The regions above (closer to the observer) and below (farther to the observer) the midplane are also labelled. The colours of each cloud refer to their LOS velocity, while their sizes reflect the weight given to each cloud towards the total emission; the larger the size of the cloud, the more this cloud contributes to the broad-line emission. The edge-on view of the BLRs of Mrk 509 shows that its asymmetry is due to a highly obscuring midplane; for PDS 456, a preference for broad-line emission to come from the far side of the BLR.}
    \label{fig:edge_only}
\end{figure*}

The inferred BLR radius for Mrk 509 is log $R_\mathrm{BLR}$ [ld] = 2.29$^{+0.01}_{-0.26}$. This is slightly larger than what is measured previously by other works \citep{Peterson1998}. However, the best-fit BH mass of Mrk 509 is $\log \rm{M}_{\rm BH}$ [$M_\odot]$ = 8.00$^{+0.06}_{-0.23}$, consistent with previous estimates \citep[e.g.,][]{Peterson1998, Grier2013}. The inferred inner BLR radius is log $R_\mathrm{BLR, min}$ [ld] = 0.93$^{+0.40}_{-0.39}$

Our BLR model for Mrk 509 prefers only $\sim$30\% of the clouds to be in circular orbits, meaning the majority ($\sim$70\%) of the clouds have significant radial (highly elliptical) motion. The resulting Bayes factor after fitting an inflowing and outflow radial model is 12.55, indicating a strong preference for the outflow radial model over inflow. This supports the result that outflows dominate the BLR of Mrk 509. 

Mrk 509 shows an asymmetric line profile with a blueshifted shoulder, and a negatively peaked differential phase signal shifted redwards of the Br$\gamma$ central wavelength (see Fig. \ref{fig:all_ave_visphi}a). Such profiles can only be produced by a BLR possessing asymmetric properties. Our best fit angular distribution parameter is $\gamma$ $\sim$ 4.1, indicating that most of the BLR clouds are on the outer faces of the BLR disc, and fewer clouds are located in the midplane of the BLR. This is physically consistent with what one would expect for an outflowing BLR. Our best fit BLR model prefers a BLR with low mid-plane transparency ($\xi$ $\sim$ 0.21) and a moderate preference for the emission to originate from the far side of the clouds ($\kappa$ $\sim$ -0.18). Fig. \ref{fig:edge_only}a shows the edge-on view of Mrk 509, emphasising the lack of clouds seen below the midplane. We note that this does not necessarily mean that there are actually fewer clouds below the midplane since we are only modelling the \emph{observed} broad-line emission. The high inclination angle allows the observer to look directly into the edge of the disc, causing the observed differential phase on the blueshifted side to be small. On the other hand, the preferential emission on the far side (where the redshifted BLR clouds are mostly found, as shown in Fig. \ref{fig:edge_only}) creates a larger phase on the redshifted side of the line centre. In addition, the midplane obscuration affects more redshifted clouds than blueshifted clouds due to the inclination angle, and therefore produces an increase on the blue wing of the flux spectrum. Therefore, the best-fit BLR model of Mrk 509 (e.g., high inclination angle, large thickness, presence of BLR asymmetry) can recreate the asymmetric flux and differential phase spectra of Mrk 509. The combination of the effects of midplane obscuration and a significant fraction of outflowing BLR clouds causes the photocentres to be curved around the centre (right column, first row in Fig. \ref{fig:photocentre_complete}), producing a strong dip in the averaged differential phase signal. This also explains why the red-blue photocentres of Mrk 509's BLR are not perpendicular to its radio jet, as the red-blue photocentres are dominated by the clouds dominated by radial motion and not by tangential/Keplerian motion.

\subsection{PDS 456}
\label{sec:disc_pds456}

Similar to Mrk 509, PDS 456 is best fit with the elliptical/radial model. This model better fits PDS 456's asymmetric differential phase spectrum, which is difficult to achieve with the circular model. The Bayesian evidence for the elliptical/radial model is $\sim$14$\times$ greater than that from the circular Pancoast model fitting. Fig. \ref{fig:corner_plot_pds456} shows the resulting corner plot of its BLR model fitting. It has a small inclination angle ($i$ $\sim$ 13$^\circ$), and the opening angle ($\theta_0$ $\sim$ 42) suggests that its BLR is moderately thick. This indicates that the system, despite the thickness, is unobscured. The BLR cloud distribution also has a steep profile ($\beta$ $\sim$ 1.83). Our BLR modelling with PDS 456 infers a BLR radius of log $R_\mathrm{BLR}$ [ld] = 2.49$^{+0.08}_{-0.38}$; the largest among our sample. The best-fit BH mass of PDS 456 is log $M_\mathrm{BH}$ [$M_\odot]$ = 8.23$^{+0.01}_{-0.49}$. Our BH mass is a factor of 10 smaller than that measured by \citet{Reeves2009} and \citet{Nardini2015} via scaling relations. This is similar to what \citet{GRAVITY2023b} concluded via dust size measurements as they also measured a smaller BH mass than previous works, albeit 0.5 dex higher than our result.bThis also suggests that the discrepancy in the BH mass between our work and that estimated by \citet{GRAVITY2023b} is just due to the assumed virial factor to calculate the latter. The inferred BLR inner radius is log $R_\mathrm{BLR, min}$ [ld] $1.16^{+0.42}_{-0.38}$.

Our modelling results with PDS 456 emphasise that almost half of the BLR system is dominated by radial motions, as the best-fit value of $f_\mathrm{ellip}$ $\sim$ 0.52. Letting $f_\mathrm{flow}$ as a free parameter causes the BLR fitting results to slightly prefer inflowing radial motions ($f_\mathrm{flow}$ $<$ 0.5). However, the Bayes evidence of fitting with $f_\mathrm{flow}$ fixed to inflow and outflow is equal to 1.818, which means that PDS 456 does not show a preference between inflow and outflow radial motions. Therefore, we conclude that the data cannot distinguish between the inflow and outflow model, and we choose to fix $f_\mathrm{flow}$ $>$ 0.5 for PDS 456 as other observations reveal outflowing signatures at both smaller and larger scales than the BLR  \citep{Reeves2003, Luminari2018, Bischetti2019}.

PDS 456 also exhibits a negative averaged differential phase spectrum almost centred at the Pa$\alpha$ central wavelength, similar to Mrk 509. Its phase signal is more asymmetric compared to that of Mrk 509, as it shows a more extended redward tail. Again, we look at the best-fit asymmetry parameters and the model photocentres to explain this phase signal. The best-fit value of $\xi$ is $\sim$ 0.75, indicating that the midplane is mostly transparent. However, the best-fit value of $\kappa$ is $\sim$ -0.44, suggesting that the BLR clouds emit preferentially from the far side. The effects of these parameters are evident in Fig. \ref{fig:edge_only}b, where the edge-on view of its BLR is shown. Here, the number of observable BLR clouds above and below the midplane are very similar, as expected for the best-fit $\xi$. However, the clouds below the midplane are displayed with larger sizes compared to those above the midplane, which reflects the best-fit $\kappa$. The preferential emission of the far side of the BLR, where the redshifted clouds reside, creates a larger phase on the redshifted side of the line centre. In addition, similar to Mrk 509, the combination of the effects of outflowing BLR clouds and their preferential emission cause the locus of model photocentres of PDS 456 to look curved (right column, second row, in Fig. \ref{fig:photocentre_complete}). The null model BLR photocentre of PDS 456 is located above the BLR photocentres, which creates the negative differential phase signal. Similar to that of Mrk 509, the dominance of radial motion in the BLR clouds causes the red-blue photocentres to be shifted less perpendicular to its radio jet. In conclusion, the cause of the asymmetric signal in PDS 456's average differential phase spectrum is due to (1) the BLR clouds' outflowing radial motions and (2) the preference of the Pa$\alpha$ emission to originate from the far side of the BLR.

\subsection{Mrk 1239}

The red- and blueshifted photocentres of Mrk 1239 show an insignificant separation (see Fig. \ref{fig:photocentre_complete}). The photocentre fitting results of Mrk 1239 suggest that the differential phase is dominated by the continuum phase, and the BLR differential phase is very moderate (low S/N). However, the perpendicular orientation of the BLR velocity gradient and continuum photocentre and the radio jet is consistent with the picture of a rotating BLR. Despite the relatively weak signal of Mrk 1239, we argue that this aids in adding further constraint to our BLR fitting compared to when we only fit its flux spectrum (as proven in Sect. \ref{sec:visphi_importance}). Given the low signal, we choose to fit the data with the simpler, circular model. The flux and differential phase spectra of Mrk 1239 are well-fit with the circular model as shown in Fig. \ref{fig:all_ave_visphi}c. While there is no clear differential phase signal for Mrk 1239, the combined fitting of the emission line profile and differential phase allows for meaningful constraints on the BLR size and SMBH mass as shown in the corner plot (see corner plot in Fig. \ref{fig:corner_plot_mrk1239}). While Fig. \ref{fig:all_ave_visphi}c does not show a clear differential phase signal, the very lack of a signal combined with the impressively low noise level still allows the model to converge and constrain the key BLR and SMBH properties of Mrk 1239.

The measured BLR PA of Mrk 1239 is $\sim$ 197$^\circ$, similar to what is measured for the red-blue photocentres of Mrk 1239, even if their separation is insignificant. The best-fitting parameters indicate that the data favours a face-on disc ($i$ $\sim$ 11$^\circ$) with a thickness of $\theta_0$ $\sim$ 42$^\circ$. This geometry is consistent with the results of \citet{Pan2021}, who suggest the observed broad emission of Mrk 1239 is actually reprocessed through scattering by polar dust. Effectively, we are observing from the LOS of the polar dust, which has a face-on viewing angle.

The cloud distribution follows an exponential profile ($\beta$ $\sim$ 1.21), while the mean radius is measured to be log $R_\mathrm{BLR}$ [ld] = 1.77$^{+0.03}_{-0.74}$. The inferred BH mass is log $M_\mathrm{BH}$ [$M_\odot]$ = 7.47$^{+0.15}_{-0.92}$. Our inferred BLR radius is consistent within uncertainties with those published previously \citep{Du2014a, GRAVITY2023b}, and so is our BH mass \citep{Kaspi2005, Du2014a, GRAVITY2023b}. Lastly, the inferred BLR inner radius is log $R_\mathrm{BLR, min}$ [ld] = $0.64^{+0.27}_{-0.28}$. However, we caution that the BLR size of Mrk 1239 is marginally constrained, as its lower error is relatively larger than its upper error. This means that the BLR size and, therefore, the BH mass of Mrk 1239 has a higher chance of being lower than our reported best-fit values.

As discussed in Sect. \ref{sec:expected_visphi}, the average observed differential phase signal of Mrk 1239 suggests that its BH mass may be similar to the \citet{Pan2021} measurement. Indeed, from the modelling, the best-fit BH mass is log $M_\mathrm{BH}$ [$M_\odot]$ = 7.47, close to the log $M_\mathrm{BH}$ [$M_\odot]$ = 7 measurement of \citet{Pan2021}.

\subsection{IC 4329A}
\label{disc_ic4329a}
Even though the noise level of individual differential phase spectra in each baseline is $\sim0.1^\circ$, the S-shape signal is not as obvious as that of Mrk 1239 (after removing the continuum phase). Nevertheless, the radio jet of IC 4329A is shown to be perpendicular to the red-blue photocentre orientation of its BLR, indicating the rotation of its BLR and we therefore fit the data with the simpler circular model. As for Mrk 1239, the combination of the emission line profile and differential phase data strongly constrains the BLR structure and SMBH mass. Similar to Mrk 1239, IC 4329A does not show a clear phase signal (Fig. \ref{fig:all_ave_visphi}d). As discussed in Sect. \ref{sec:visphi_importance}, this lack of clear phase signal acts as a strong constraining factor in IC 4329A's BLR and SMBH properties. Without the differential phase spectrum, the BLR properties of IC 4329A will not be well constrained.

The cloud distribution of its BLR ($\beta$ $\sim$ 1.81) suggests it has a steep inner radial profile. The inclination angle is $i$ $\approx$ 54$^\circ$; similar to \citet{Bentz2023}, the high inclination angle of IC 4329A suggests that the AGN system and its host galaxy disc are misaligned significantly. The opening angle is $\theta_0$ $\approx$ 54$^\circ$, indicating that the BLR is thick. We resolved the BLR of IC 4329A, with a BLR radius of log $R_\mathrm{BLR}$ [ld] = 1.13$^{+0.10}_{-0.23}$ and a BH mass of log $M_\mathrm{BH}$ [$M_\odot]$ = 7.15$^{+0.38}_{-0.26}$. The inferred BLR inner radius is log $R_\mathrm{BLR, min}$ [ld] = $0.64^{+0.27}_{-0.28}$.

\citet{Bentz2023} recently published results of a new RM campaign for IC 4329A which spanned about 5 months during 2022. They used \texttt{CARAMEL} \citep{Bentz2021b, Bentz2022} to fit their RM data with the BLR model from \citet{Pancoast2014}, which is similar to what we use. Our best-fit BLR size, BH mass, $i$, and $\theta_0$ are all consistent with RM measurements from \citet{Bentz2023}. Indeed, the average observed phase signal of IC 4329A spans the same range as the expected phase signal of IC 4329A with its BH mass fixed to what \citet{Bentz2023} measured.

\section{The GRAVITY-AGN radius-luminosity relation}
\label{sec:R-L}

One of the main objectives of our GRAVITY-AGN large programme is to place AGN properties derived by GRAVITY in the context of AGN scaling relations previously derived by other methods, namely the R-L and $M_\mathrm{BH}$-$\sigma_*$ relations. In this section, we show the first non-RM-derived R-L relation. To start, we have the BLR radius and BH mass measurements of our 4 targets, and adding the measurements of the 3 previously published targets (IRAS 09149-6206 from \citealt{GRAVITY2020a}, 3C 273 from \citealt{GRAVITY2018}, and NGC 3783 from \citealt{GRAVITY2021a}), gives us 7 AGNs to plot in the R-L space, now spanning a wide range of luminosity (log $\lambda L_{\lambda} (5100 \AA)$ $\sim$ 43.0 - 46.5). We fit these data points with a power-law relation similar to that of \citet{Bentz2013}:
\begin{equation}
    {\rm log} \ (R_{\rm BLR}/ {\rm ld}) = K + \alpha \ {\rm log}(\lambda L_\lambda/10^{44} \ {\rm erg} \ {\rm s}^{-1}).
\end{equation}
We employed the \texttt{LINMIX} algorithm \citep{Kelly2007}, a package that utilises a hierarchical Bayesian approach to linear regression with measurement errors and a parameter, $\sigma^2$, which encapsulates the intrinsic random scatter around the regression. We calculate the 16th, 50th, and 84th percentile of the resulting posterior distributions of each parameter to get their best-fit values and their 1$\sigma$ uncertainties. We use the optical AGN luminosities of our targets listed in Table \ref{tab:targets} and the best-fit BLR sizes of our targets to produce our GRAVITY-AGN R-L relation. However, we emphasise that Mrk 1239 is known to be a physically complicated object based on various observed properties (e.g., Balmer decrement, polarisation) leading to the high obscuration of its nucleus \citep{Goodrich1989, Doi2015}. Its observed broad lines and optical continuum are also suggested to originate from scattering by polar dust around the BLR \citep{Pan2021}. Therefore, we cannot directly use the observed $\lambda L_{\lambda} (5100 \AA)$ of Mrk 1239. We opt to use the extinction-free $\lambda L_{\lambda} (5100 \AA)$ of Mrk 1239 from \citet{Pan2021}.

\begin{figure}
    \centering
    \includegraphics[width=0.45\textwidth]{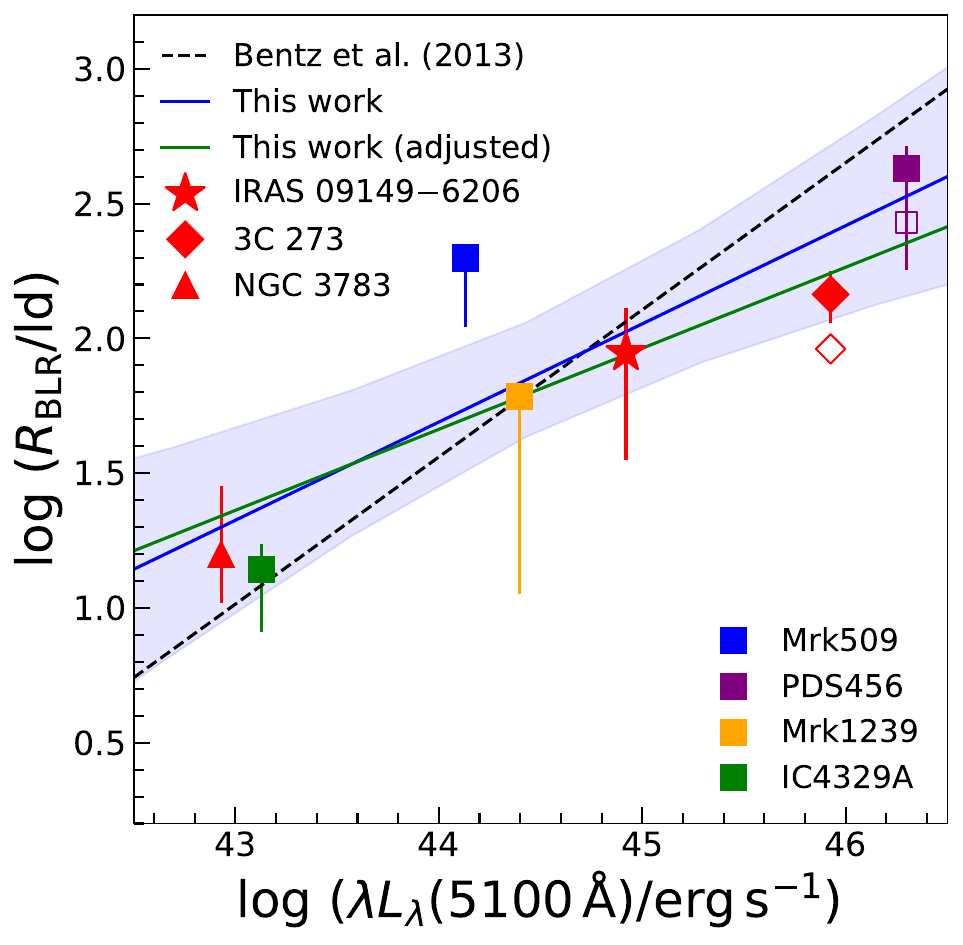}
    \caption{Logarithm of the BLR radius as a function of log $\lambda L_{\lambda} (5100 \AA)$ (R-L relation). The blue line shows the best-fit line derived from all 7 GRAVITY-observed AGNs, and the blue shaded region is its 1$\sigma$ confidence interval. The AGN luminosity is taken from Table \ref{tab:targets}. The dashed line represents the best-fit radius luminosity relation from \citet{Bentz2013}. The 3 published AGNs (IRAS 09149-6206 from \citealt{GRAVITY2020a}, 3C 273 from \citealt{GRAVITY2018}, and NGC 3783 from \citealt{GRAVITY2021a}) are shown as a red star, red diamond, and red square, respectively. Our 4 targets are shown in differently coloured filled squares with 1$\sigma$ errors. Our best-fit R-L (power-law) relation has a slope of $\alpha$ = 0.37$^{+0.18}_{-0.17}$, which is lower but, given the current large uncertainty, still consistent within uncertainties with the slope of R-L relation fit from \citet{Bentz2013} ($\alpha$ = 0.533$^{+0.035}_{-0.033}$). If we follow the prescription from photoionisation models, the BLR sizes of 3C 273 and PDS 456 will be adjusted (shown as open symbols), and the fitted GRAVITY R-L relation is shown as the green line.}
    \label{fig:rl_rel}
\end{figure}

Fig. \ref{fig:rl_rel} shows the result of our R-L relation fit compared with that of \citet{Bentz2013}. The best-fit values for our GRAVITY-observed AGNs are $K$ = 1.69$^{+0.23}_{-0.23}$, $\alpha$ = 0.37$^{+0.18}_{-0.17}$, and $\sigma^2$ = 0.23$^{+0.48}_{-0.13}$. The large uncertainties in our best-fit parameters are inevitable due to our small sample size. By observing more AGNs with GRAVITY, we can increase the sample size and, therefore, better constrain the R-L relation independent of RM (see Sect. \ref{sec:conclusion} for more discussion of these future prospects). We emphasise that our approach allows us to extend the R-L relation to much higher luminosities without the need for decade-long time baselines to observe their expected year-long time lags. Our results suggest that the GRAVITY R-L relation has a flatter slope ($\sim$ 0.4) compared to that of the canonical R-L relation from \citet{Bentz2013}, which has a slope of $\sim$ 0.5. \citet{Woo2023} recently showed a similar conclusion as this work by focussing on reverberation mapping of high-luminosity AGNs (10$^{44}$ L$_\odot$ $<$ $\lambda L_{\lambda} (5100 \AA)$ $<$ 10$^{45}$ L$_\odot$, with a few sources possessing $\lambda L_{\lambda} (5100 \AA)$ $>$ 10$^{45}$ L$_\odot$).

We also consider the relative time lags of the emission lines with respect to the H$\beta$ time lag based on the predictions of photoionisation models. These models \citep[e.g.,][]{Netzer2020, Korista2004} predict that the BLR size is dependent on certain parameters such as the optical depth of the observed emission line and the variability and the photoionisation flux of the central engine. \citet{Kuhn2023} recently confirmed the consistency of observed relative time lags of several emission lines (e.g., H$\alpha$, He {\tiny I}) with respect to H$\beta$ time lag with what the radiation pressure confined (RPC) BLR model by \citet{Netzer2020} described. According to this model, the time lags of Br$\gamma$ and Pa$\alpha$ are about 1.0-1.2 and 1.5-1.7 times larger than that of H$\beta$. It is clear, therefore, that BLR sizes derived from Pa$\alpha$ are greatly affected, and thus, an adjustment must be taken into consideration. We show the adjusted BLR sizes of 3C 273 and PDS 456 (the objects with Pa$\alpha$ line profiles) as open symbols in Fig.~\ref{fig:rl_rel}. If we consider these adjusted BLR sizes in our GRAVITY R-L relation fit, we get $K$ = 1.66$^{+0.24}_{-0.24}$, $\alpha$ = 0.30$^{+0.19}_{-0.19}$, and $\sigma^2$ = 0.27$^{+0.54}_{-0.16}$,. We, therefore, conclude that the resulting slope of the R-L relation will be shallower if we consider prescriptions from photoionisation models and adjust BLR sizes measured from Pa$\alpha$ lines. As for the effect of this "BLR stratification" phenomenon on the rest of the BLR model parameters, future work \citep{Kuhn2023} on NGC 3783 has shown that the geometry beyond the radial distribution of the line emission (e.g., inclination angle, opening angle, black hole mass) are consistent among the different emission lines (H$\alpha$, H$\beta$, Pa$\alpha$, Pa$\beta$, HeI).

\subsection{Effect of Eddington ratio in our results}

\begin{table*}
\caption{Eddington ratio ($\lambda_{\rm Edd}$ = $L_\mathrm{bol}$/$L_\mathrm{Edd}$ and dimensionless accretion rate $\dot{\mathcal{M}}$ from Eqn. 4 of \citealt{Du2018}) of all seven targets calculated from GRAVITY observations and by other literature. The $L_\mathrm{bol, use}$ for each target is also presented.}
\label{tab:edd_ratio}     
\centering                         
\begin{tabular}{ccccc}       
\hline              
\hline
Object & \makecell{log $L_{\rm bol}$ \\ (erg s$^{-1}$)} & \makecell{$\lambda_{\rm Edd}$ \\ (GRAVITY)} & \makecell{$\lambda_{\rm Edd}$ \\ (other works)} & \makecell{$\dot{\mathcal{M}}$ \\ (GRAVITY)}\\
\hline
Mrk 509 & 45.32$^{a}$ & 0.16$^{1}$ & 0.19$^{5}$ & 1.9 \\
PDS 456 & 47.01$^{b}$ & 4.64$^{1}$ & 0.44$^{6}$ & 2.0 $\times 10^2$ \\
Mrk 1239 & 45.36$^{c}$ & 0.60$^{1}$ & 0.10 - 2.8$^{7}$ & 13 \\ 
IC 4329A & 45.12$^{a}$ & 0.72$^{1}$ & 0.13 - 0.46$^{8,9}$ & 1.4 \\
\hline
3C 273 & 46.64$^{b}$ & 1$^{2}$ & 0.6$^{10}$ & 22 \\
NGC 3783 & 44.52$^{a}$ & 0.05$^{3}$ & 0.06$^{11}$ & 0.02 \\
IRAS 09149-6206 & 45.29$^{a}$ & 0.1$^{4}$ & 0.4$^{12}$ & 5 \\
\hline
\end{tabular}
\vspace{0.1mm}
\\
{\footnotesize \begin{itemize}
\item[$^a$] log $L_{\rm bol, 14-195 keV}$ from \citet{Koss2017} and \citet{Baumgartner2013}.
\item[$^{b}$] Based on converting the log $\lambda L_\lambda (5100 \AA)$ to log $L_{\rm bol, 5100 \AA}$ using the prescription from \citet{Trakhtenbrot2017}.
\item[$^{c}$] log $L_{\rm bol, 5100 \AA}$ calculated from \citet{Pan2021}
\end{itemize}
\begin{flushleft}
\textbf{References:} 
(1) This work, 
(2) \cite{GRAVITY2018},
(3) \cite{GRAVITY2021a}, 
(4) \cite{GRAVITY2020a}, 
(5) \cite{Fischer2015},
(6) Calculated using the BH mass from \citet{Nardini2015} and the bolometric luminosity from \citet{Reeves2000},
(7) \cite{Jiang2021},
(8) \cite{Ogawa2019},
(9) \cite{delaCallePerez2010}
(10) \cite{Husemann2013},
(11) \cite{Brenneman2011},
(12) \cite{Walton2020}
\end{flushleft}
}
\end{table*}

An important aspect to discuss with the GRAVITY R-L relation fit is the Eddington ratio of our targets since previous works have suggested that it is the "third" parameter in the R-L relation, and it can explain the smaller observed BLR sizes compared to what the canonical R-L relation expects \citep{Du2018, Du2019} especially for PDS 456 whose best-fit BH mass and BLR size are much smaller than previous estimates. In this work, we calculate the Eddington ratio as $\lambda_\mathrm{Edd}$ = $L_\mathrm{bol}$/$L_\mathrm{Edd}$, where $L_\mathrm{Edd}$ = 1.3 $\times$ 10$^{38}$ ($M_\mathrm{BH}$/$M_\odot$) erg s$^{-1}$. We also calculate the dimensionless accretion rate, $\dot{\mathcal{M}}$, based on the standard disc model \citep{Shakura1973} shown in Eqn. 4 of \citet{Du2018}. We calculate the former to easily facilitate comparison with other literature, while we calculate the latter to compare with the super-Eddington accreting massive black hole (SEAMBH) sample of \citet{Du2018}. We use the best-fit inclination angle of each object to calculate $\dot{\mathcal{M}}$. However, due to the presence of outflowing radial motion in Mrk 509 and PDS 456, their best-fit inclination angles may not be easily as straightforward as the rest of the GRAVITY-observed AGNs, and therefore, the values of their $\dot{\mathcal{M}}$ should be taken with caution.

For Mrk 509, IC 4329A, NGC 3783, and IRAS 09149-6206, we use the 14-195 keV luminosity from the BAT AGN Spectroscopic Survey \citep{Koss2017}, which is directly taken from the 70-month Swift-BAT survey \citep{Baumgartner2013}, to calculate $L_\mathrm{bol}$ through the calibration in \citep{Winter2012}. For 3C 273 and PDS 456, we decided to use the bolometric luminosity calculated from the 5100 $\AA$ luminosity and the bolometric correction of \citet{Trakhtenbrot2017}, which was also used by \citet{GRAVITY2020c}. This is because 3C 273 is most likely dominated by the jet in the X-ray regime \citep{Dermer1997, Vasudevan2007}, and PDS 456 exhibits strong X-ray variability, causing the observed $L_\mathrm{2-10 \ keV}$ to be 0.2\% of its bolometric luminosity \citep{Reeves2009}. We also use the 5100 $\AA$ luminosity and bolometric correction of \citet{Trakhtenbrot2017} to get the bolometric luminosity of Mrk 1239 from the $L_{\lambda} (5100 \AA)$ measurement of \citet{Pan2021}. This bolometric luminosity is comparable to (only slightly larger than) the bolometric luminosity estimated from $WISE$ W3 (12 $\mu$m). Both of these $L_\mathrm{bol}$ values are much larger than the bolometric luminosity estimated from the 2-10 keV X-ray luminosity \citep{Jiang2021}, which is likely because of the high obscuration that is not accounted for.

Table \ref{tab:edd_ratio} lists the calculated $\lambda_\mathrm{Edd}$ and $\dot{\mathcal{M}}$ of our 7 targets based on GRAVITY-derived properties, together with the $L_\mathrm{bol}$ used to calculate $\lambda_{\rm Edd}$ in this work. The $\lambda_\mathrm{Edd}$ values calculated from GRAVITY-derived values are consistent with those taken from the literature, except PDS 456. As for $\dot{\mathcal{M}}$, we see that the range of the $\dot{\mathcal{M}}$ of our targets are similar to that of \citet{Du2018} (about 10$^{-3}$ to 10$^{3}$). PDS 456 shows a relatively large Eddington ratio compared to the previous estimate assuming a bolometric luminosity of $\sim$ 10$^{47}$ erg s$^{-1}$ \citep{Reeves2000} and a BH mass of log $M_\mathrm{BH}/M_\odot$ = 9.24 \citep{Nardini2015}. However, \citet{Yang2021} also find evidence that the accretion rate of PDS 456 may exceed the Eddington rate based on the observed X-ray wind velocities on PDS 456.


Our Eddington ratio calculations suggest that the BLR sizes of our targets are consistent with the picture that the accretion rate plays a role in the observed deviation from the canonical R-L relation, as shown by \citet{Du2018} and \citet{Du2019}. These previous works have shown that $\mathcal{R}$(Fe {\tiny II}), which is thought to be an indicator of accretion rate, is correlated with the deviation of AGNs from the canonical R-L relation. However, recent results from \citet{Woo2023} undermine these conclusions, as they found the aforementioned correlation to be weaker after adding more luminous sub-Eddington AGNs. Nevertheless, our sample size is limited, and therefore, more AGN observations with GRAVITY will help provide independent evidence to further test this hypothesis.

\subsection{Effect of used luminosity in the R-L relation}

We also investigate what happens to the R-L relation if we use a different indicator of AGN luminosity other than $\lambda L_\lambda (5100 \AA)$, the standard luminosity for reporting the R-L relation. It is possible $\lambda L_\lambda (5100 \AA)$ is not a good measure of a target's ionising luminosity at high luminosities if, for example, the SED shape changes significantly. In this case, the bolometric luminosity (measured from X-ray observations) might be considered. If we use the log $L_\mathrm{bol}$ listed in Table. \ref{tab:edd_ratio} to plot the R-L relation, we get a best-fit slope of $\alpha$ = 0.49$^{+0.30}_{-0.30}$ (in our discussion, we disregard the effect of relative time lags with respect to H$\beta$ for simplicity). This suggests that the apparent decrease in the slope of the R-L relation might instead be caused by a non-linear UV-to-optical relationship. Similarly, \citet{Netzer2019} presents a luminosity-dependent bolometric correction. Using this to convert our log $\lambda L_\lambda (5100 \AA)$ values to $L_\mathrm{bol}$, we find the slope of the R-$L_{\mathrm AGN}$ relation to be 0.48$^{+0.22}_{-0.21}$.  Therefore, while we are limited to our small sample size, our large luminosity range has revealed that a changing SED shape could instead explain the apparent departure of the R-L relation from R $\sim$ $L^{0.5}$. This will be further assessed with a much larger sample of GRAVITY-observed AGNs.

\citet{Woo2023} also recently investigated the possibility of log $\lambda L_\lambda (5100 \AA)$ not being a good representative of the AGN luminosity of their targets. Their tests show the systematic change of the SED slope between the UV and optical wavelengths may be partly responsible for the deviation from the 0.5 slope of the R-L relation, similar to what our test above suggests.

\section{The black hole mass - stellar velocity dispersion relation}
\label{sec:M-sigma}

\begin{figure}
    \centering
    \includegraphics[width=0.45\textwidth]{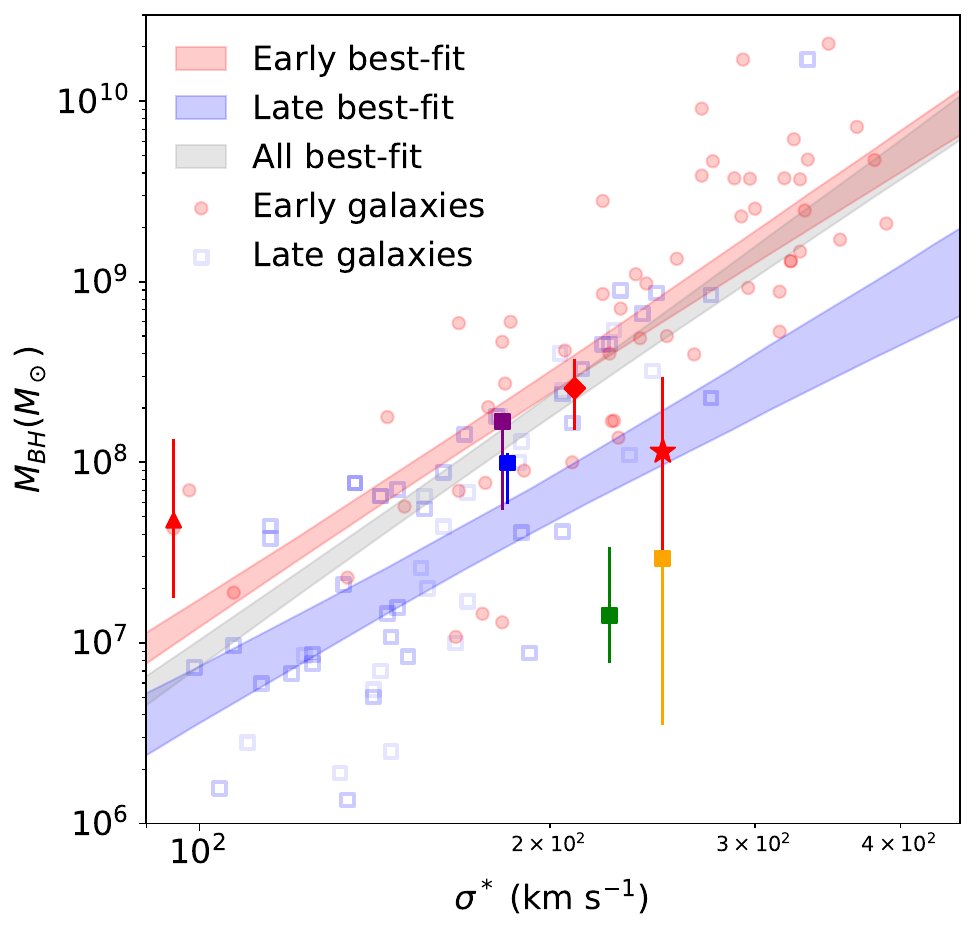}
    \caption{Plot showing BH mass vs stellar velocity dispersion ($M_\mathrm{BH}$-$\sigma_*$ relation). The symbols of our 7 AGNs are similar to that of Fig. \ref{fig:rl_rel}. The figure is redrawn from Fig. 3 of \citet{Greene2020}, but with the addition of the 4 targets introduced in this work. Early- and late-type galaxies are shown as red circles and blue hollow squares, respectively, while BH mass upper limits of some low-mass late-type galaxies are shown as inverted blue triangles. All these galaxies' measurements are taken from \citet{Greene2020} and referenced therein. The best-fit $M_\mathrm{BH}$-$\sigma_*$ relation for early- and late-type galaxies are shown as red and blue-shaded regions, while the best-fit relation for all galaxies is shown as a grey-shaded region. The best-fit lines are surrounded with their 1$\sigma$ confidence interval, and all best-fit values are also taken from \citet{Greene2020}.
    }
    \label{fig:m-sigma}
\end{figure}

Since one of the ultimate goals of our work is to acquire precise BH masses, we are also interested in where our targets lie in the $M_\mathrm{BH}$-$\sigma_*$ relation. Fig. \ref{fig:m-sigma} compares our data points with $M_\mathrm{BH}$-$\sigma_*$ relations from other studies. We focus our comparison with \citet{Greene2020}, wherein they fit the $M_\mathrm{BH}$-$\sigma_*$ relations for early- and late-type galaxies. This sample comprises various subsamples of galaxies from different works \citep[e.g.,][]{Kormendy2013, Greene2016, Saglia2016, Krajnovic2018, Thater2019, denBrok2015, Nguyen2018, Nguyen2019}. Upper limits of galaxies with smaller stellar velocity dispersions were estimated via stellar dynamics (see Table 4 of \citealt{Greene2020} and references therein). The stellar velocity dispersions of our targets are drawn from the literature. For Mrk 509, Mrk 1239, and IC 4329A, we adopt the measurements from \citet{Oliva1999} derived from Si 1.59 $\mu$m, CO (6,3) 1.62 $\mu$m, and CO(2,0) 2.29 $\mu$m. We use the average $\sigma_*$ if there are two or more available measurements for each object, and the typical errors of their measured $\sigma_*$ are about $\pm$20 km/s. For PDS 456, we use the calculated dynamical mass from \citet{Bischetti2019} to derive its tentative stellar velocity dispersion. They acquired CO(3-2) emission maps of PDS 456, from which they measured the dynamical mass of PDS 456 within 1.3 kpc. The inferred dynamical mass is 1.0 $\times$ 10$^{10}$ $M_\odot$. We can estimate the stellar velocity dispersion via virial theorem: $\sigma_*$ = $\sqrt{\frac{GM}{R}}$, where $G$ is the gravitational constant, $M$ is the dynamical mass measured at $R$ = 1.3 kpc. With this method, we estimate $\sigma_*$ $\sim$ 182 km/s for PDS 456. However, we caution readers that this estimate is uncertain, as an unknown geometric correction factor $C$ is usually added (especially if the object is not a disc) \citep{Binney2008}. In a variety of galactic mass distributions, C = 6.7 could be an appropriate value to assume \citep[e.g.,][]{FoersterSchreiber2009}. In addition, the most accurate way of measuring the stellar velocity dispersion of a target is through measuring equivalent widths of stellar absorption lines \citep[e.g.,][]{Oliva1999}. However, this is yet to be done for PDS 456. We also caution that the $\sigma_*$ of IRAS 09149-6206 used in this work is based on the [O{\small III}] line and therefore is very uncertain \citep{GRAVITY2020a}.

All of our objects lie close to the standard relation, showing similar scatter to other samples. However, Mrk 1239 and IC 4329A are shown to be below the local $M_\mathrm{BH}$-$\sigma_*$ relation. As shown in Table \ref{tab:edd_ratio}, these two objects also have relatively high Eddington ratios. These results are consistent with previous works showing highly accreting BHs to be below the local BH-galaxy scaling relations \citep{Ding2022, Zhuang2023}. However, we find 3C 273 and PDS 456, the sources with the highest Eddington ratios from our sample, to be placed consistently with the $M_\mathrm{BH}$-$\sigma_*$ relation of early-type galaxies and above the $M_\mathrm{BH}$-$\sigma_*$ relation of late-type galaxies. We note that the $\sigma_*$ of PDS 456 is very uncertain, but it is not the case for 3C 273. Nevertheless, the two sources are the most luminous AGNs at low redshift, which by selection requires a high BH mass and high accretion rate.

\section{Possible origin of the spatial BLR offset and its relation with AGN luminosity}
\label{sec:blr_offset_vs_lum}

\begin{figure*}
    \centering
    \includegraphics[width=\textwidth]{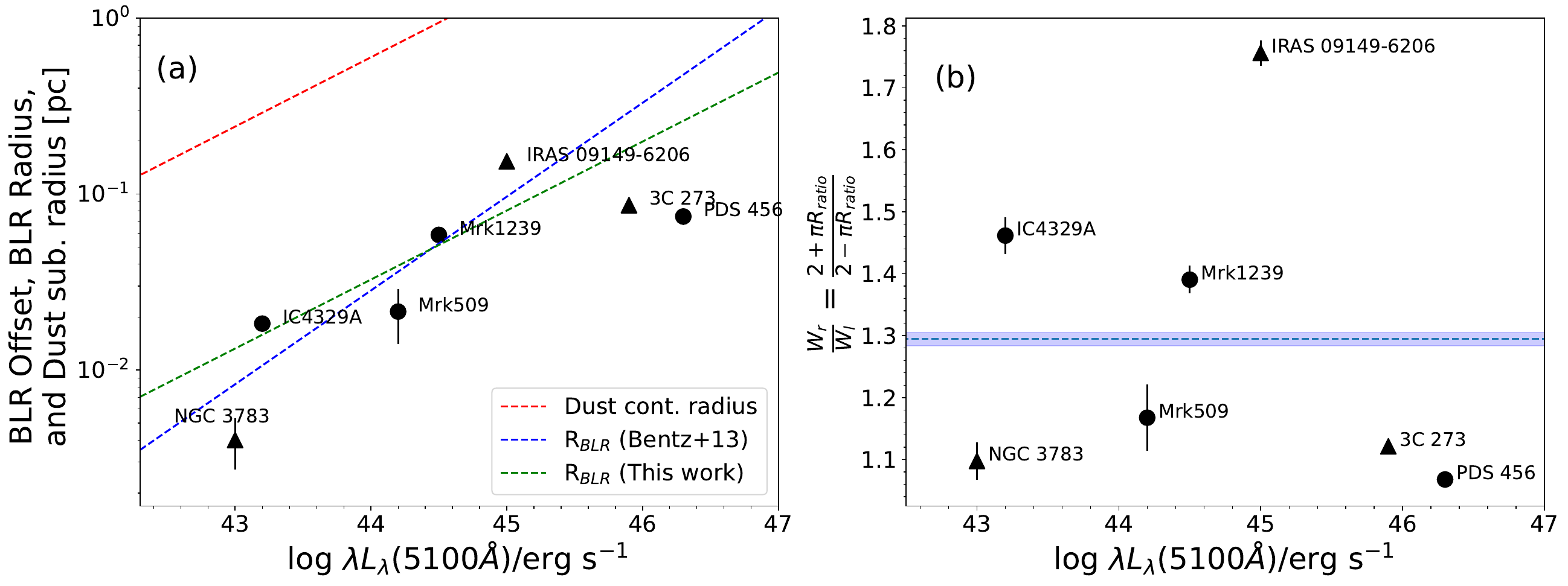}
    \caption{Plots showing the spatial offset between the BLR and continuum photocentres and its possible physical representation and their connection with optical AGN luminosity. (a) The BLR offset and (b) the flux ratio of two sides of the hot dust ($W_\mathrm{r}$/$W_\mathrm{l}$) as a function of optical AGN luminosity, assuming that the hot dust emits asymmetrically as explained in Sect. \ref{sec:blr_offset_vs_lum}. The 4 targets introduced in this work are shown in circles, while the previously published AGNs (3C 273, IRAS 09149-6206, and NGC 3783) are shown in triangles. The error bars are 1$\sigma$ errors calculated through the propagation of uncertainties from the centroid position of the null photocentre fitting. For comparison, the relation between hot dust size and AGN luminosity \citep{GRAVITY2023b} is shown by the red dashed line, while the relation between the BLR radius and AGN luminosity (the so-called R-L relation) based on \citet{Bentz2013} is shown by the blue dashed line. We also show the fitted R-L relation in this work depicted by the green dashed line. The Pearson correlation coefficients and p-values for the left and right plots are 0.81 and 0.026, and -0.15 and 0.73, respectively.}
    \label{fig:offset_flux_ratio_vs_lum}
\end{figure*}

Previously in Sect. \ref{sec:photocentre}, we find a spatial offset between the BLR and the continuum photocentre for all of our targets. We call this the "BLR offset", and it ranges from $\sim$30-140 $\mu$as for our four targets. Upon investigation, we find a strong positive (Pearson correlation coefficient = 0.81) correlation (p = 0.026) between the BLR offsets (henceforth called $R_\mathrm{off}$) and optical AGN luminosity of all GRAVITY-observed AGNs. We show this in Fig. \ref{fig:offset_flux_ratio_vs_lum}a, where we also compare the data with R-L relation taken from \citet{Bentz2013}, the fitted R-L relation in this work, and the dust continuum R-L relation derived from our GRAVITY data \citep{GRAVITY2023b}. Our comparison with these relations suggests two things about the BLR offsets: (1) that they are all within the dust sublimation radius and (2) that they are of the same scale as BLR sizes. Due to the latter, we can rule out the BLR structure as a possible origin of the observed BLR offsets.

We include the BLR offsets of the 3 previously published GRAVITY AGNs in Fig. \ref{fig:offset_flux_ratio_vs_lum}a. The photocentre of IRAS 09149-6206 is reported in \citet{GRAVITY2020a}. We newly measured the photocentre offsets of 3C 273 and NGC 3783. For 3C 273, the reduction method in \citet{GRAVITY2018} removed the continuum phase signal, so we reduce the data again with our new method based on that of \citet{GRAVITY2020a}. For NGC 3783, we already reduce the data with our new method, so we simply perform photocentre and BLR fitting again\footnote{The BLR offset reported in Table 2 of \citet{GRAVITY2021a} has the "secondary" component in its differential phase removed. This component is due to an offset hot dust $\sim$0.6 pc away from the main central hot dust component. Our new photocentre fitting does not remove this component.}. We measure a new BLR offset of 22.8 $\pm$ 2.1 $\mu$as and 71.6 $\pm$ 6.2 $\mu$as for 3C 273 and NGC 3783, respectively. This does not affect the derived BLR differential phase of the two targets.

\begin{figure}
    \centering
    \includegraphics[width=0.45\textwidth]{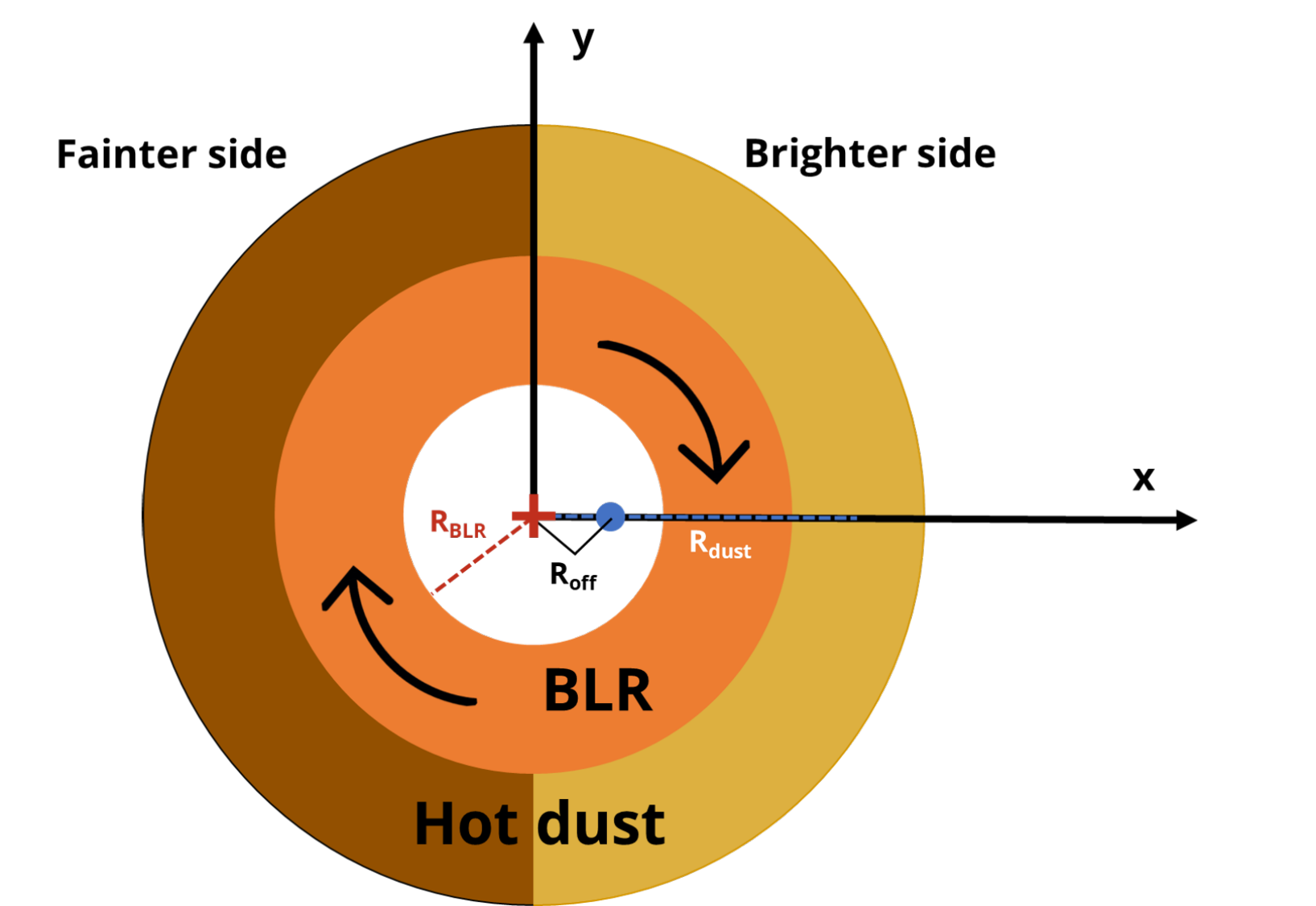}
    \caption{2D schematic diagram showing our explanation for the observed offset between the BLR and hot dust. The BLR (orange disc) is rotating, has a size of $R_\mathrm{BLR}$, and its photocentre is marked as a red cross and is assumed to be the same as the position of the central BH. For simplicity, the hot dust (with a radius of R$_{\rm dust}$) is assumed to have two sides: the left side being the fainter one and the right side being the brighter one. This causes the continuum photocentre (blue dot) to shift to the brighter side of the hot dust. The observed offset between the BLR and hot dust photocentres is labelled R$_{\rm off}$.
    }
    \label{fig:model}
\end{figure}

Given the luminosity dependence of the offsets, we propose that the offset between the BLR and continuum photocentres results from asymmetric $K$-band emission from the hot dust. This asymmetry can then be simply modelled as the hot dust having a side with a brighter flux and a side with a fainter flux. Fig. \ref{fig:model} shows the schematic diagram for an easy visualisation of the model. This asymmetry can be produced by several factors, for example, the presence of a parcel of dusty cloud/s with significant flux located within the hot dust (similar to that of \citealt{GRAVITY2021a}) or the dusty clouds are optically thick at $\sim$2 $\mu$m, so one preferentially sees the side illuminated by the AGN. Some of the emitting hot dust structures of AGNs could also have irregularities or clumpy regions, or the edge of a foreground dust lane could coincide with the LOS to the nucleus \citep{GRAVITY2020a}. If the asymmetric $K$-band emission from the hot dust is caused by the coincidence of the LOS to a BLR irregularity, a connection between the BLR offsets and inclination angles might be implied. However, the Pearson correlation coefficient and p-value of the log $R_\mathrm{BLR, off}$ [pc] and $i$ [$^\circ$] are -0.42 and 0.35, respectively, indicating the absence of a significant correlation between the two. Even if we exclude Mrk 509 and PDS 456, the targets whose best-fit inclination angles may not represent their "true" inclination angle due to the presence of outflowing radial motion in their BLRs, the resulting Pearson correlation coefficient and p-value are -0.39 and 0.51, respectively. Therefore, the asymmetric $K$-band emission from the hot dust could not be due to the coincidence of the LOS to any BLR irregularities.

\textcolor{red}{The next objective is to determine how bright (in terms of flux) the brighter side of the hot dust is relative to the fainter side.} Following our model, we investigate this by deriving a formula based on the concept of the centre of mass that will give us the flux ratio between the two sides. We arrive at the following relation (see Appendix \ref{sec:appendixB} for more discussion):
\begin{equation}
    \frac{W_r}{W_l} = \frac{2+\pi R_{\rm ratio}}{2 - \pi R_{\rm ratio}}
\end{equation}
where $R_\mathrm{ratio} = \frac{R_\mathrm{off}}{R_\mathrm{dust}}$, $R_\mathrm{dust}$ is the dust sublimation radius derived from the dust radius - luminosity relation from \citet{GRAVITY2023b}, $W$ is the flux, and the subscripts $l$ and $r$ refer to the left and right sides of the hot dust, respectively.

We show $W_\mathrm{r}$/$W_\mathrm{l}$ as a function of optical AGN luminosity in Fig. \ref{fig:offset_flux_ratio_vs_lum}b. The average value of $W_\mathrm{r}/W_\mathrm{l}$ is 1.29 $\pm$ 0.01, meaning that the brighter side of the hot dust has a $\sim$ 30\% higher flux than its fainter side. The Pearson correlation p-value of $W_\mathrm{r}$/$W_\mathrm{l}$ vs. log $\lambda L_\lambda (5100 \AA)$ is $\sim$ 0.73, suggesting insignificant correlation between the two quantities.

The absence of a significant relationship between the BLR offsets and inclination angle also purports that the "tilted torus" model \citep{Lawrence2010} could not explain the presence of BLR offsets. Other possible reasons include variable obscuration on the BLR and smaller scales. \citet{Dehghnian2019} proposed that a "holiday" period (a period when the covering factor of the LOS obscurer varies) can explain the decorrelation of the emission lines and continuum variations observed on NGC 5548. However, it would be too much of a coincidence to conclude that \textit{all} the targets were observed during their "holiday" periods so that their BLR photocentres are offset with respect to their continuum photocentres.

\section{Virial factors}
\label{sec:f}

Finally, we calculate the virial factors of our 4 targets. \citet{DallaBonta2020} argued that the dispersion or $\sigma$ (i.e., square root of the second moment of the line profile) is better than the FWHM in calculating the virial factor due to the latter introducing a bias in the BH mass scale. \citet{Yu2019} also purported a similar conclusion as they found the $\sigma$ is insensitive to the inclination angle, and hence, the BLR geometry doesn't affect the resulting BH mass. Hence, we follow this prescription and measure the $\sigma$ of the model line profile. Afterwards, the values of the relevant model parameters are drawn randomly from the sampled posterior parameter space created during BLR model fitting. We use the formula $f_{\sigma} = GM_\mathrm{BH} / (R_\mathrm{BLR} v^2_{\sigma})$ to calculate the virial factor.

\begin{figure}
    \centering
    \includegraphics[width=0.45\textwidth]{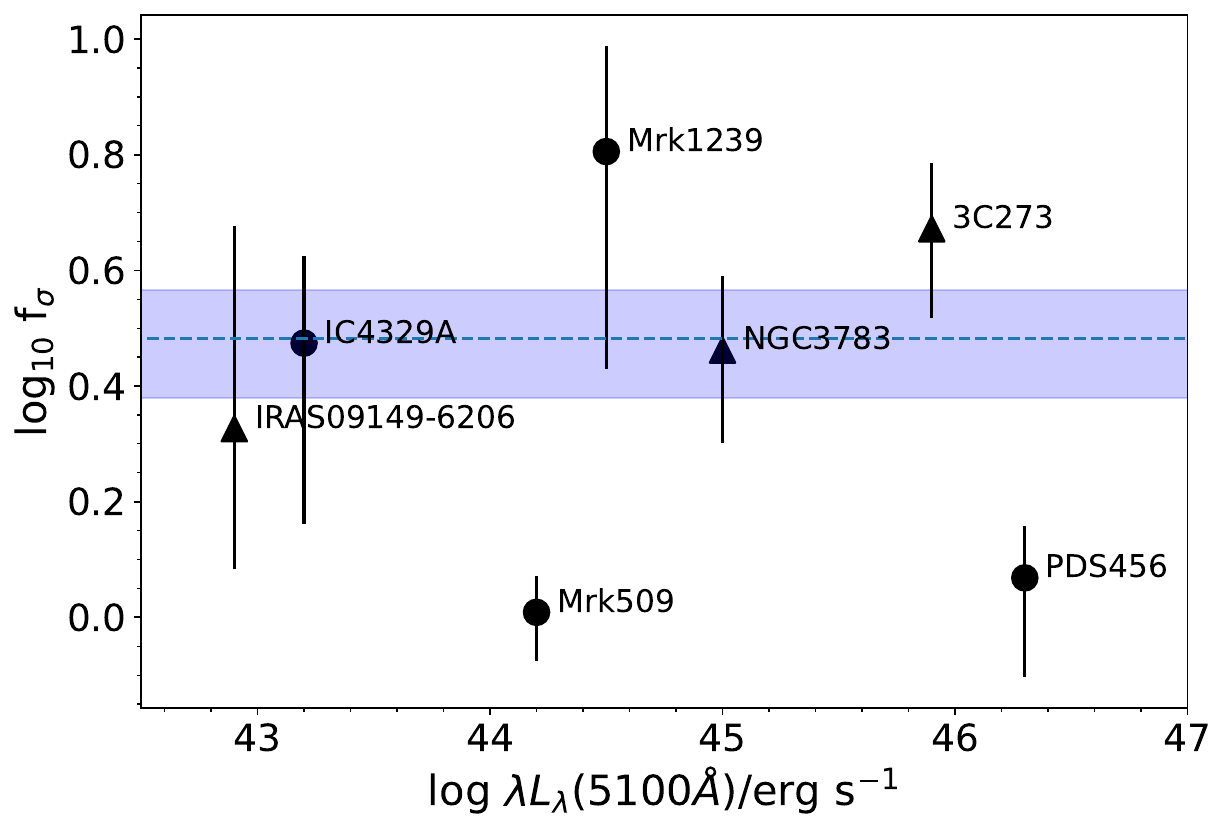}
    \caption{Virial factor ($f_{\sigma}$) as a function of optical AGN luminosity. The legends are similar to that of Fig. \ref{fig:offset_flux_ratio_vs_lum}. The Pearson correlation p-value is $\sim$ 1.0, indicating an insignificant correlation between the two quantities.}
    \label{fig:virial_factors}
\end{figure}

Fig. \ref{fig:virial_factors} shows the virial factors of our sources as a function of optical AGN luminosity. The virial factors of Mrk 509, PDS 456, Mrk 1239, and IC 4329A are 1.10$^{+0.16}_{-0.18}$, 1.17$^{+0.27}_{-0.38}$, 6.39$^{+3.32}_{-3.37}$, and 2.98$^{+1.23}_{-1.53}$, respectively. The virial factor of 3C 273 was taken from \citet{GRAVITY2018}, while the virial factors of NGC 3783 and IRAS 09149-6206 are calculated from the posterior distributions of their parameters. The average virial factor of all 7 AGNs is $\langle f_{\sigma} \rangle$ = 3.04 $\pm$ 0.64. This is lower than that calculated by previous works that assumed their AGN sample follows the same $M_\mathrm{BH}$-$\sigma_*$ as those of quiescent galaxies \citep{Onken2004, Woo2010, Park2012, Batiste2017}, which is $\langle f_{\sigma} \rangle$ $\sim$ 5. It is not clear whether we should expect a match here, since the host galaxy type plays a role in the determination of the $M_\mathrm{BH}$-$\sigma_*$ relation. And, our average virial factor is consistent with that of \citet{Grier2017a} and \citet{Williams2018}, who also calculated the individual virial factors of their AGN sample and got $\langle f_{\sigma} \rangle$ $\sim$ 3. We also note that we find lower $f_\sigma$ values for Mrk 509 and PDS 456, which intriguingly are the two targets with significant radial motion in their BLR.

The resulting Pearson correlation coefficient between log ${\rm f}_{\sigma}$ and log $\lambda L_{\lambda} (5100 \AA)$ is $\sim$ -0.02, and the p-value is $\sim$ 0.96, indicating insignificant correlation. Previous works \citep{Villafana2023, Williams2018} reported an positive correlation, although non-significant (p-values $>>$ 0.05), between the two quantities. They also reported a possible positive correlation between log $f_\sigma$ and log $M_\mathrm{BH}$/$M_{\odot}$. However, calculating the Pearson correlation coefficient and p-value of these two quantities in our work suggests otherwise (-0.27 and 0.56, respectively). Hence, we conclude that there is no correlation between the virial factor and optical luminosity or BH mass for our targets. However, the uncertainties are large, and bigger samples will be needed to confirm this result.

\section{Conclusions and future prospects}
\label{sec:conclusion}

We investigate the broad Br$\gamma$ line-emitting regions of Mrk 509, Mrk 1239, and IC 4329A, and the broad Pa$\alpha$ line-emitting region of PDS 456. To study the kinematics and properties of their BLRs, we performed photocentre and BLR model fitting. Our results support many of the assumptions and scaling relations used and derived by RM, and they provide an independent method that can be extended to high-$z$ and large BH mass:

\begin{enumerate}
    \item Most of the AGNs observed by GRAVITY can be well described by a thick, rotating disc of clouds, consistent with RM assumptions. However, two of our targets (Mrk 509 and PDS 456) show evidence of outflowing BLR clouds, while the other two targets (Mrk 1239 and IC 4329A) have relatively weaker differential phase signals that limit the constraints of their BLR sizes and BH masses. Nevertheless, we prove from our analyses that the differential phase is crucial in providing stronger constraints on the BLR kinematics.

    \item By adding the other three AGNs that were previously observed with GRAVITY (3C 273, IRAS 09149-6206, and NGC 3783), we derived a new R-L relation based on GRAVITY-derived BLR sizes only. We derived a slope and intercept of $\alpha$ = 0.37$^{+0.18}_{-0.17}$ and $K$ = 1.69$^{+0.23}_{-0.23}$, respectively. Our results are consistent with works showing shallower R-L relation slopes, although the effect of the Eddington ratio in this relation cannot be fully realised with our relatively small sample.

    \item Most of the GRAVITY AGNs are consistent within the scatter of the standard $M_\mathrm{BH}$-$\sigma_*$ relation. Mrk 1239 and IC 4329A, two of our targets with relatively high Eddington ratios, are placed below the $M_\mathrm{BH}$-$\sigma_*$ relation consistent with previous works, while 3C 273 and PDS 456, the targets possessing the highest Eddington ratios among our sources, are still consistent within the $M_\mathrm{BH}$-$\sigma_*$ relation, suggesting that their high luminosity may play a role in their difference compared to other high Eddington-ratio sources such as Mrk 1239 and IC 4329A.

    \item We find a significant correlation between the offset between the BLR and continuum photocentre and optical AGN luminosity, and the offsets of all GRAVITY AGNs are found to be of a similar scale as their BLR sizes. This correlation is surmised to be due to asymmetric $K$-band emission of the hot dust, and a simple model was created to estimate that this emission is $\sim$30\% brighter on one side than the other for our sample.

    \item Lastly, we calculated the virial factors of our four targets and found an average virial factor (based on the dispersion of the model line profiles) of $\langle f_{\sigma} \rangle$ = 3.04 $\pm$ 0.64 for all GRAVITY-observed AGNs. This is consistent with previous works that calculated the individual virial factors of their targets.

Further increasing the observed AGNs with GRAVITY will greatly help in providing better hints in answering some of the remaining questions unanswered by this work. With the advent of GRAVITY+, the limits of GRAVITY will be pushed even further, providing wide-field off-axis fringe-tracking, newer adaptive optics systems on all the UTs, and laser guide stars. With GRAVITY+, observing hundreds of AGNs and even those at high redshift will be achievable \citep{GRAVITY2022}. An example of this feat is the first dynamical mass measurement of a $z\sim$2 quasar with GRAVITY-Wide \citep{GRAVITY2023b}. It is expected, therefore, that we may also be able to shed light on the R-L relation at high redshift and provide more accurate BH mass measurements for galaxies at the cosmic noon.

\end{enumerate}

\begin{acknowledgements}
    This research has used the NASA/IPAC Extragalactic Database (NED), operated by the California Institute of Technology, under contract with the National Aeronautics and Space Administration. This research has used the SIMBAD database, operated at CDS, Strasbourg, France.
\end{acknowledgements}

\bibliography{main}{}
\bibliographystyle{aa}

\appendix
\section{Corner plots of BLR model fits}
\begin{figure*}
    \centering
    \includegraphics[width=\textwidth]{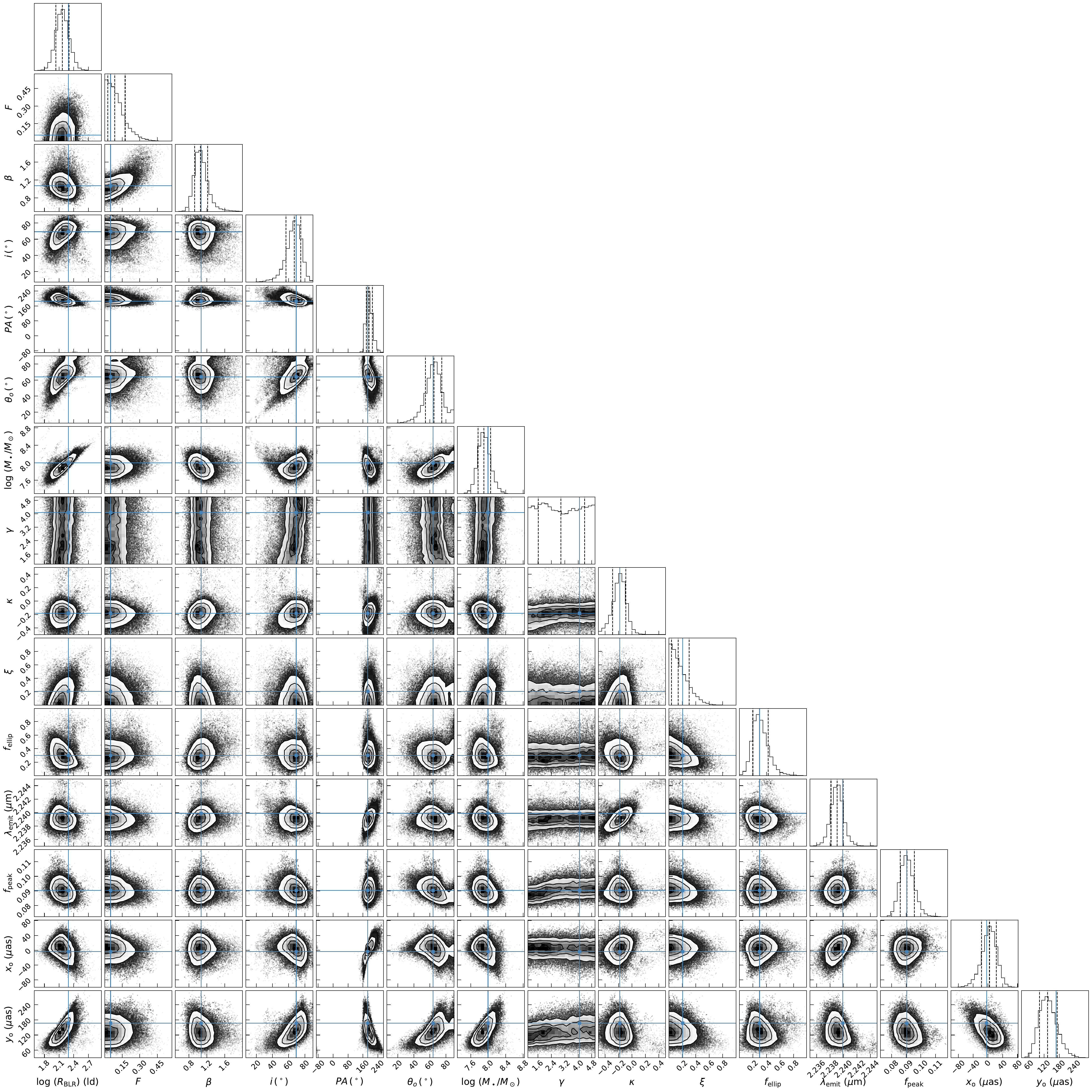}
    \caption{Posterior probability distribution of the fitted parameters from the elliptical and radial model fitting Mrk 509 data. The blue lines and cross points refer to the maximum a posteriori value. The vertical dashed lines represent the 68\% (1$\sigma$) credible intervals.}
    \label{fig:corner_plot_mrk509}
\end{figure*}

\begin{figure*}
    \centering
    \includegraphics[width=\textwidth]{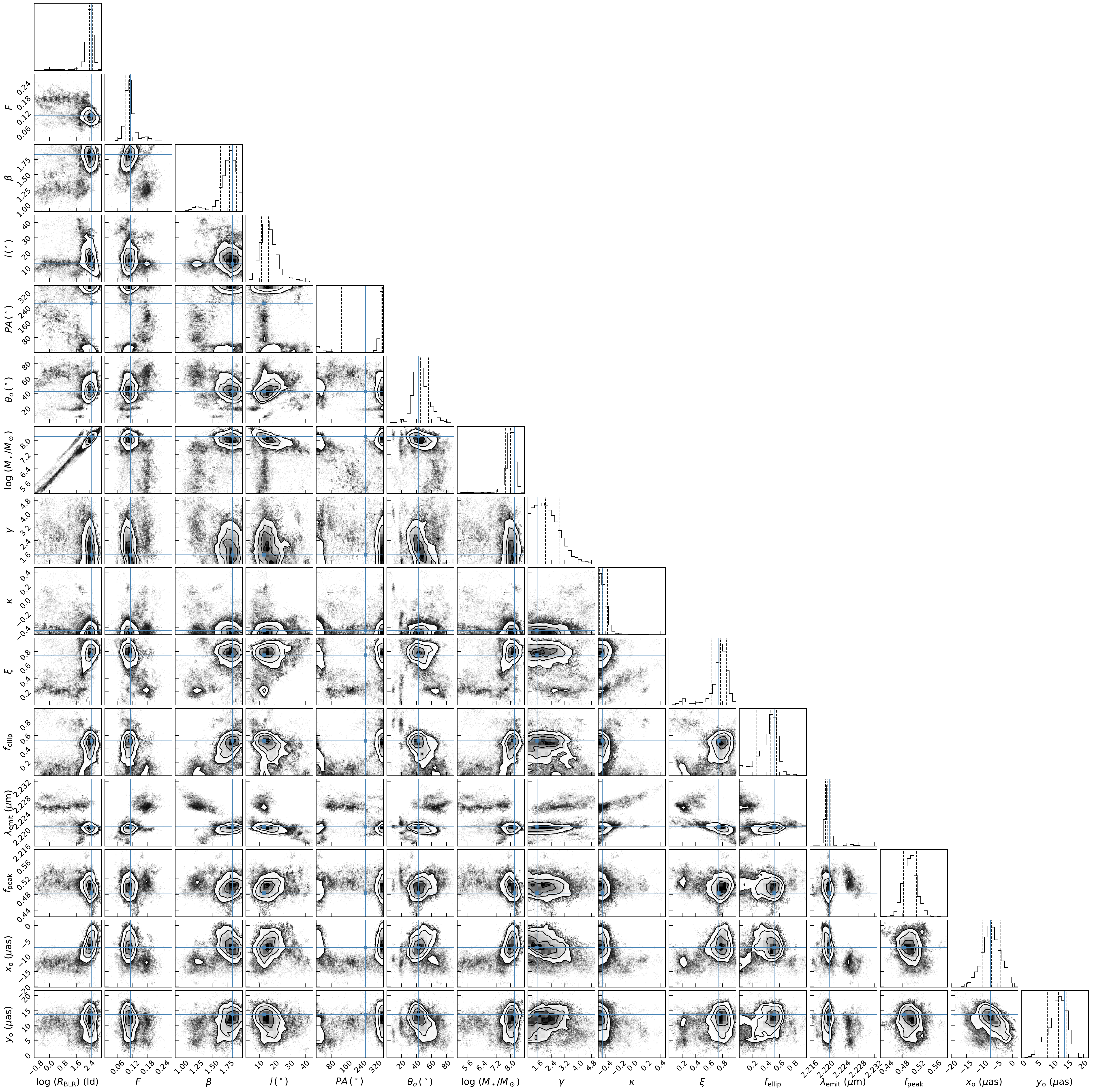}
    \caption{Similar to Fig. \ref{fig:corner_plot_mrk509} but for PDS 456.}
    \label{fig:corner_plot_pds456}
\end{figure*}

\begin{figure*}
    \centering
    \includegraphics[width=\textwidth]{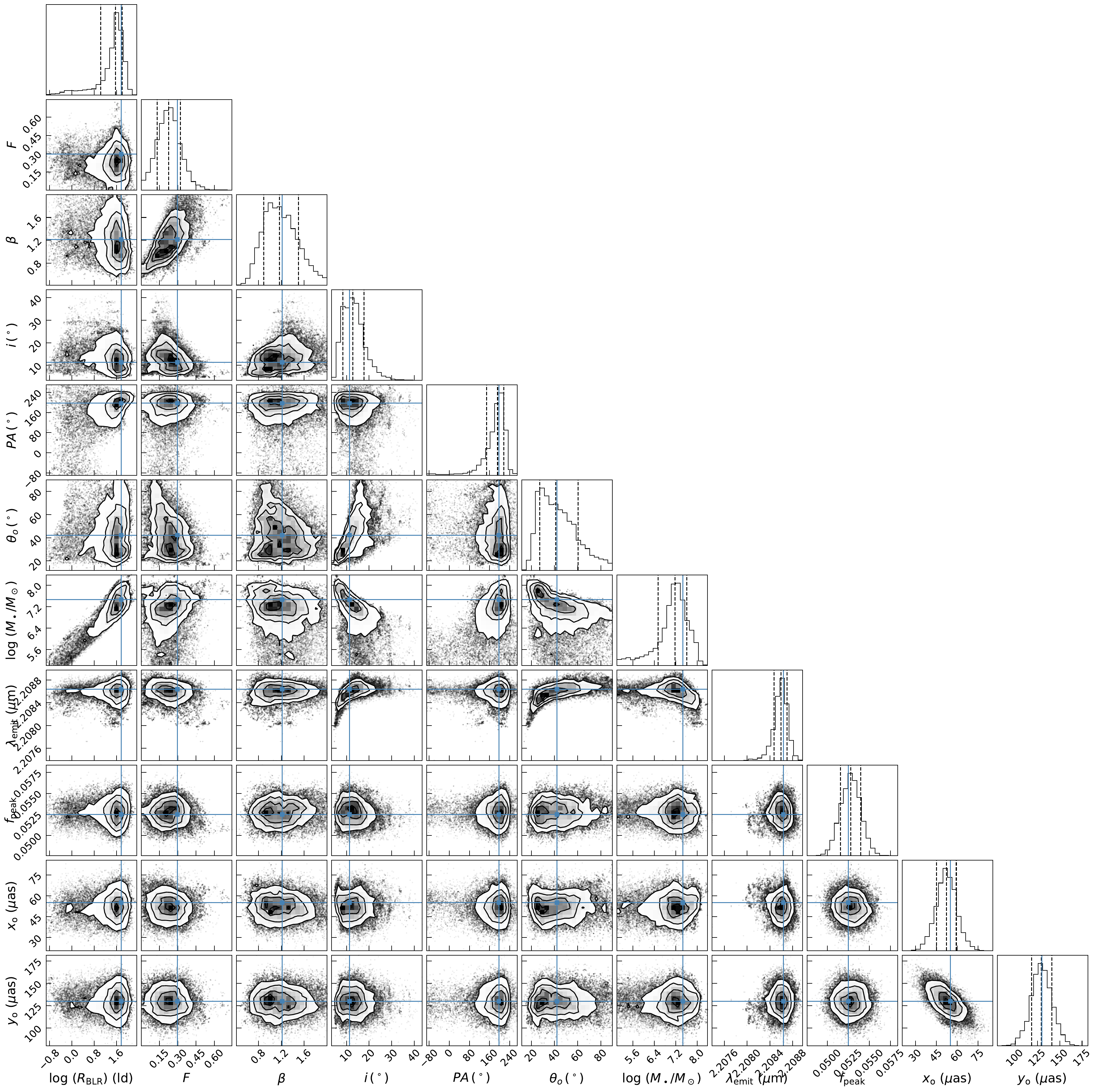}
    \caption{Posterior probability distribution of the fitted parameters from the circular model fitting Mrk1239 data. The legends from Fig. \ref{fig:corner_plot_mrk509} still apply.}
    \label{fig:corner_plot_mrk1239}
\end{figure*}

\begin{figure*}
    \centering
    \includegraphics[width=\textwidth]{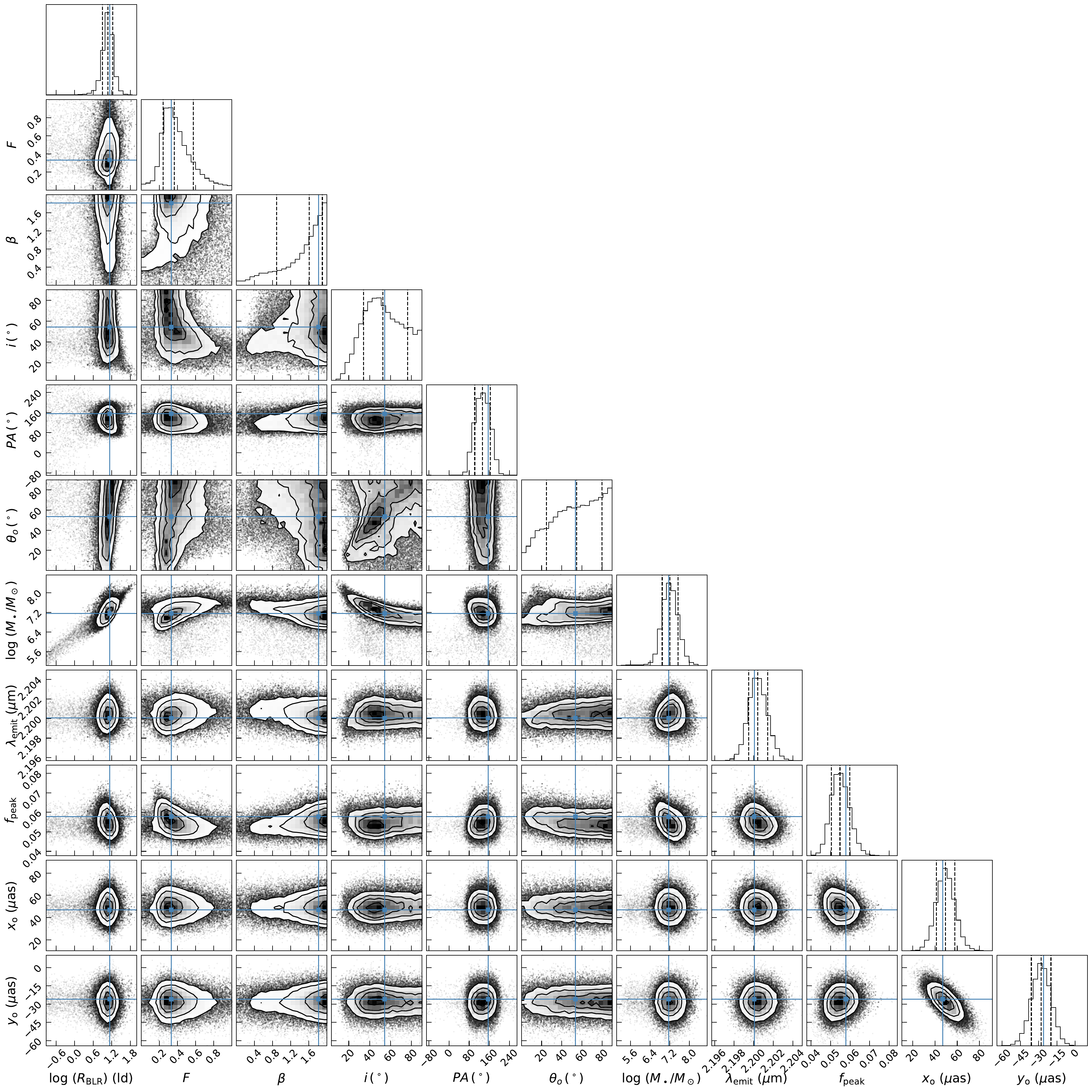}
    \caption{Similar to Fig. \ref{fig:corner_plot_mrk1239} but for IC 4329A.}
    \label{fig:corner_plot_ic4329a}
\end{figure*}

The posterior probability density distributions of the fitted parameters for all of our targets are shown in Figs. \ref{fig:corner_plot_mrk509} to \ref{fig:corner_plot_ic4329a} as corner plots.

\section{Derivation of the flux ratio of two sides of the hot dust}
\label{sec:appendixB}

First, let $P_\mathrm{cont}$ be the distance of the hot dust photocentre to the origin:
\begin{equation}
    P_{\rm cont} = \frac{\Sigma w_i r_i}{\Sigma w_i}
\end{equation}
where $w_\mathrm{i}$ is the weight/flux of a dust clump at $r_\mathrm{i}$. We define $P_\mathrm{BLR}$ as the photocentre position of the BLR. In our simple model, the central BH is located at the origin of our coordinate system, and $P_\mathrm{BLR}$ is also situated at the origin. Considering the left and right side of the BLR, we have
\begin{equation}
    P_{\rm cont}^{\prime} = \frac{\Sigma w_l r_l + \Sigma w_r r_r}{\Sigma w_l + \Sigma w_r}
\end{equation}
where the subscripts $l$ and $r$ refer to the left and right sides of the hot dust, respectively. The summations must be converted into definite integrals with the range of angles encompassing the left and right sides as their respective limits:
\begin{equation}
    \begin{array}{l}
        P_{\rm cont} = \frac{1}{\int_{\pi/2}^{3\pi/2} w_l \mathrm{d}\theta +  \int_{-\pi/2}^{\pi/2} w_r \mathrm{d}\theta} ( {W}_{\rm l} R_{\rm dust}\int_{\pi/2}^{3\pi/2} \cos\theta \ \mathrm{d}\theta \\
        + {W}_{\rm r} R_{\rm dust} \int_{-\pi/2}^{\pi/2} \cos\theta \ \mathrm{d}\theta)
    \end{array}
\end{equation}
where $W_\mathrm{l}$ and $W_\mathrm{r}$ are the total flux of the left and right sides of the hot dust, respectively. $R_\mathrm{dust}$ is the average size of the hot dust (we disregard the effect of a possible large discrepancy between the inner and outer radius of the hot dust ring for simplicity). The equation simplifies to:
\begin{equation}
    P_{\rm cont} = \frac{2R_{\rm dust}(W_r - W_l)}{\pi(W_l + W_r)}.
\end{equation}
If we re-define $R_\mathrm{off}$ = $P_\mathrm{cont}$ (since $P_\mathrm {BLR}$ coincides with the origin and the continuum photocentre simply represents the actual offset), we get $\frac{R_\mathrm{off}}{R_\mathrm{dust}}$ = $\frac{2}{\pi} \frac{W_\mathrm{r} - W_\mathrm{l}}{W_\mathrm{r} + W_\mathrm{l}}$. To calculate $W_\mathrm{r}/W_\mathrm{l}$, we have:

\begin{equation}
    \frac{W_r}{W_l} = \frac{2+\pi R_{\rm ratio}}{2 - \pi R_{\rm ratio}}
\end{equation}
\begin{flushleft}
where $R_\mathrm{ratio} = \frac{R_\mathrm{off}}{R_\mathrm{dust}}$.
\end{flushleft}

\end{document}